\pdfoutput=1 
\documentclass[iop]{emulateapj}

\shorttitle{Resolving the Radio Source Background}
\shortauthors{Condon et al.}

\begin{document}


\title{Resolving the Radio Source Background: \\
    Deeper Understanding Through Confusion}


\author{J.~J.~Condon\altaffilmark{1}, W.~D.~Cotton\altaffilmark{1}, 
E.~B.~Fomalont\altaffilmark{1}, and K.~I.~Kellermann\altaffilmark{1}}
\affil{National Radio Astronomy Observatory, 520 Edgemont Road,
Charlottesville, VA 22903, USA}
%%\email{jcondon@nrao.edu}

\author{N.~Miller}
\affil{Department of Astronomy, University of Maryland, 
College Park, MD 20742-2421, USA}

\author{R.~A.~Perley\altaffilmark{1}}
\affil{National Radio Astronomy Observatory, P.O. Box 0,
Socorro, NM 87801, USA}

\author{D.~Scott, T.~Vernstrom, and J.~V.~Wall}
\affil{Department of Physics and Astronomy, University of British Columbia,
Vancouver, BC V6T\,1C1, Canada}


\altaffiltext{1}{The NRAO is a facility of the National Science Foundation
operated by Associated Universities, Inc.}


\begin{abstract}
We used the Karl G.~Jansky Very Large Array (VLA) to image one primary
beam area at 3~GHz with $8''$ FWHM resolution and $ 1.0 \,\mu{\rm
  Jy~beam}^{-1}$ rms noise near the pointing center.  The $P(D)$
distribution from the central 10 arcmin of this confusion-limited
image constrains the count of discrete sources in the $1 < S(\mu{\rm
  Jy}) < 10$ range.  At this level the brightness-weighted
differential count $S^2 n(S)$ is converging rapidly, as predicted by
evolutionary models in which the faintest radio sources are
star-forming galaxies; and $\approx 96$\% of the background
originating in galaxies has been resolved into discrete sources.
About 63\% of the radio background is produced by AGNs, and the
remaining 37\% comes from star-forming galaxies that obey the
far-infrared (FIR) / radio correlation and account for most of the FIR
background at $\lambda \approx 160\,\mu$m.  Our new data confirm that
radio sources powered by AGNs and star formation evolve at about the
same rate, a result consistent with AGN feedback and the 
%% begin revision, inserted because of referee's ``main concern''
rough
%% end revision correlation of black hole and bulge stellar masses.
The confusion at centimeter wavelengths is low enough that neither the
planned SKA nor its pathfinder ASKAP EMU survey should be confusion
limited, and the ultimate source detection limit imposed by
``natural'' confusion is $ \leq 0.01\,\mu$Jy at $\nu = 1.4$~GHz.  If
discrete sources dominate the bright extragalactic background reported
by ARCADE\,2 at 3.3~GHz, they cannot be located in or near galaxies
and most are $\leq 0.03\,\mu$Jy at 1.4~GHz.
\end{abstract}





\section{INTRODUCTION}

\subsection{Counts of Extragalactic Radio Sources}

Together the number per steradian $n(S)dS$ of discrete extragalactic radio
sources having flux densities $S$ to $S + dS$ and the brightness
temperature $T_{\rm b}$ that sources contribute to the sky background
constrain the nature and evolution of all extragalactic radio
sources, even those too faint to be counted individually. 
%% Begin Jasper Wall's historical intro 
Counts of radio sources achieved instant fame and notoriety with the
announcement by \citet{ryl55a} and \citet{ryl55b} that the source count from
the 2C survey was dramatically steeper than that expected for a
uniformly-filled Euclidean universe.  The remarkable \citet{ryl55b}
paper claimed that the majority of ``radio stars'' were extragalactic,
that they were so distant and radio-luminous as to be mostly beyond
the reach of then-existing optical telescopes, and that the steep count
showed dramatic cosmic evolution for individual objects in space
density or luminosity. Then revolutionary, it remains a succinct
summary of our current understanding of extragalactic radio sources.

The claimed evolution sparked a bitter and public dispute, refuting as
it did the popular steady-state cosmology \citep{bon48,hoy48}. Also
the 2C radio source count was made before radio astronomers
appreciated the importance of confusion and Eddington bias.
\citet{mil57} showed the great majority of the 2C sources were just
``confusion bumps,'' the broad 2C beam often blending two or more
faint background sources to appear as a single stronger
source. \citet{mil57} also noted that the count from their
higher-resolution survey showed ``no cosmological effects'' with the
possible exception of mild clustering.  By that time, however, the
Cambridge group had fully understood what went wrong, and the
aftermath triggered (1) the 3C and 4C Cambridge surveys which were
carefully cognizant of confusion, (2) the pioneering $P(D)$ analysis
of confusion by \citet{sch57} which showed how to extract the true
source count from a confusion-limited survey, and (3) the first
aperture-synthesis images \citep{ryl62} which revealed a decreased
source-count slope (loosely termed ``turnover'') at flux densities
below 1 Jy.

By 1966 there was general agreement about the counts of sources
stronger than $S \sim 0.1$~Jy at low frequencies. The 4C confusion
$P(D)$ \citep{hew61} and direct \citep{gow66} counts had mapped their
main features: a rise steeper than the static Euclidean slope followed
by a decrease to sub-Euclidean slopes at flux densities below 1
Jy. Independent sky surveys, such as the Parkes survey \citep{bol64},
were in agreement. To explain the source count, several investigators
developed mathematical (non-physical) models for the cosmic evolution
of extragalactic radio sources. \citet{lon66} first showed that the
relatively rapid change of the count slope could be explained by
differential cosmic evolution, in which the more luminous sources
underwent more evolution in their comoving space density than their
less-luminous counterparts did.

Sensitive surveys made with the WSRT and VLA in the 1980s extended the
1.4~GHz source count down to mJy levels and below. \citet{con84c} and
\citet{win85} discovered a point of inflection near 1 mJy below which
the count slope flattens, suggesting the emergence of a new population
of radio sources.  Spiral galaxies dominate the local radio luminosity
function below $L \sim 10^{23}{\rm ~W~Hz}^{-1}$.  Radio sources
powered by star formation in spiral galaxies can account for this new
population and for nearly half of the extragalactic sky background if
they evolve at about the same rate as the stronger sources powered by
active galactic nuclei (AGNs) in elliptical galaxies
\citep{con84a}. Later models \citep{wal97b,boy98,wil08} find evolution
similar to that of the cosmic star-formation rate shown in the
Lilly-Madau diagram [see \citet{hop07} for a modern
  data-set]. 
%% begin revision: added at request of referee's ``nitpick a''
\citet{mas10} accounted for the inflection with an
evolutionary model using the empirical FIR/radio correaltion, FIR
counts, and the redshift distributions of star-forming galaxies .  
%% end revision
A recent compilation of source counts from surveys at frequaencies 150
MHz to 15 GHz is presented by \citet{dez10}; features and cosmic
implications are discussed there in detail.

 Of particular note in the present context are persistent
 discrepancies among the most sensitive 1.4~GHz direct counts that far
 exceed both Poisson and clustering uncertainties \citep{con07,owe08}.
 These discrepancies may result from different authors making
 different corrections for the effects of partial resolution.
 Resolution corrections can be large and difficult to estimate for an
 image whose resolution approaches the median angular size $\langle
 \theta_{\rm s} \rangle \sim 1''$ of faint sources.  The deep synthesis
 counts were also extended by confusion $P(D)$ distributions in
 low-resolution images \citep{mit85,win93}.  We used their approach
 here to count sources a factor of ten fainter in a low-resolution
 ($8''$ FWHM) 3~GHz image that does not need significant corrections for
 partial resolution.

\subsection{The Source Contribution Sky Brightness}

The radio source count and the source contribution to the sky brightness at
frequency $\nu$ are connected by the Rayleigh-Jeans approximation
\begin{equation}\label{dseq}
S n(S) dS = {2 k_{\rm B} d T_{\rm b} \nu^2 \over c^2}~,
\end{equation}
where $k_{\rm B}\approx 1.38 \times 10^{-23} {\rm ~J~K}^{-1}$ is the
Boltzmann constant and $dT_{\rm b}$ is the sky brightness
temperature added
by the $n(S)$ sources per steradian with flux densities $S$ to $S + dS$.  
The observed source count now spans eight decades of flux density, so it
must be plotted with a logarithmic abscissa.  Substituting $d[\ln(S)] = dS /
S$
gives the brightness temperature contribution per decade of flux density
\begin{equation}\label{dtbeq}
\biggl[{d T_{\rm b} \over d \log(S)}\biggr] = 
\biggl[{\ln(10) c^2 \over 2 k_{\rm B} \nu^2}\biggr] 
S^2 n(S)~.
\end{equation}

The cumulative brightness temperature from all sources stronger than
any detection limit $S_0$ is
\begin{equation}\label{tbcumeq}
\Delta T_{\rm b} (>S_0) = \biggl[{\ln(10) c^2 \over 2 k_{\rm B}\nu^2}\biggr] 
\int_{S_0}^\infty S^2 n(S) d[\log(S)]~.
\end{equation}
If $S_0$ is low enough, $\Delta T_{\rm b}$ approaches the total
$T_{\rm b}$ of the source background, and we can say that the
background has been resolved into known sources.  For example,
\citet{gle10} resolved about half of the far-infrared (FIR) background
at $\lambda = 250,~350$, and $500\,\mu{\rm m}$; and the Herschel/PEP
(PACS Evolutionary Probe) survey \citep{ber11} resolved about 3/4 of
the COBE/DIRBE background at $\lambda = 160\,\mu{\rm m}$.

\begin{figure}
\vskip .5in
\epsscale{1.}
\plotone{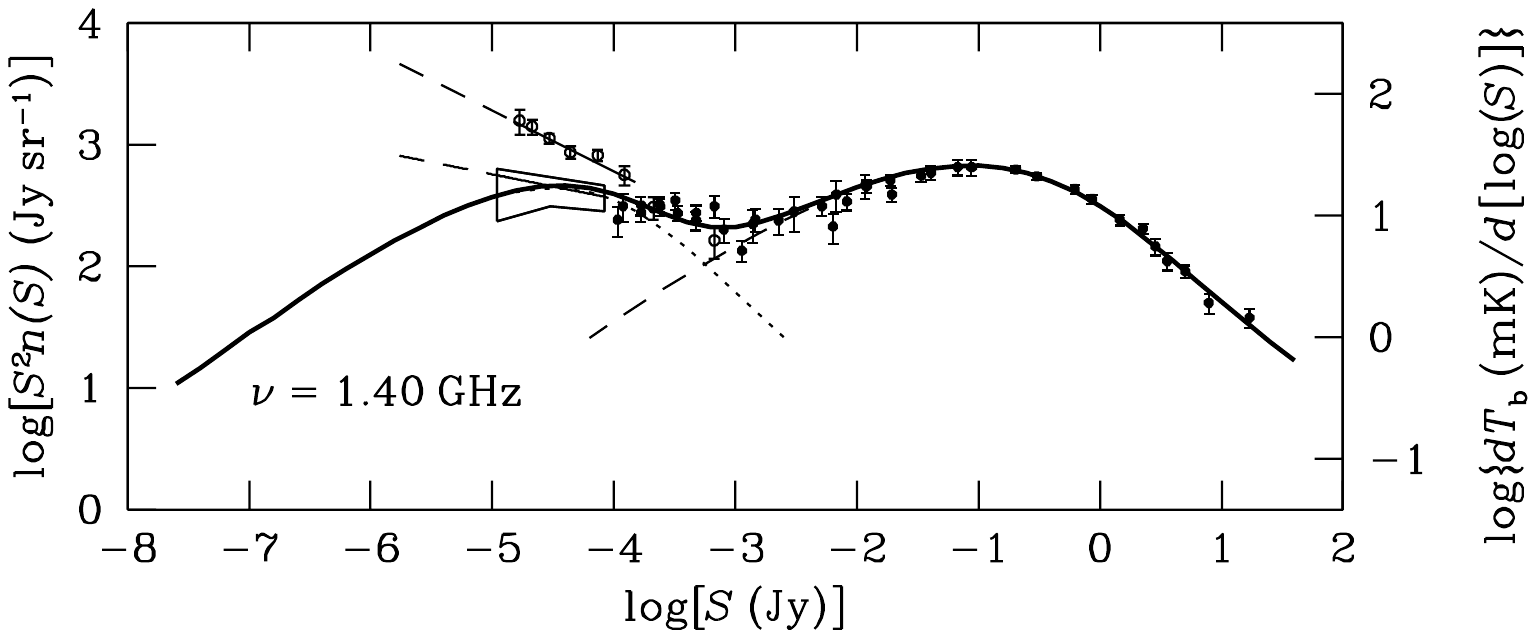}
\vskip -3.5in
\caption{Published 1.4~GHz source counts.  The brightness-weighted
  source count $S^2 n(S)$ is proportional to the contribution per
  decade of flux density to the sky background temperature $T_{\rm
    b}$.  The filled points at $\log[S\,{\rm (Jy)}] > -3$ are from
  \citet{con84a} and \citet{mit85}. The polygon encloses the range of
  1.4~GHz counts consistent with confusion \citep{mit85}, and the
  straight line inside the box is the best power-law fit to the
  confusion data.  The open data points and their power-law fit (upper
  straight line) indicate the \citet{owe08} source count.  The solid
  curve is the \citet{con84b} model count composed of sources powered
  primarily by AGNs (dashed curve) and by star formation (dotted
  curve). Left ordinate: log of the 1.4 GHz source count $S^2 n(S)$
  (Jy~sr$^{-1}$).  Right ordinate: log of the source contribution
  $dT_{\rm b} / d[\log(S)]$ (mK) to the 1.4~GHz background per decade
  of flux density.}

\label{oldtbkndfig}
\end{figure}

Figure~\ref{oldtbkndfig} shows the flux-density range covered by
published 1.4~GHz source counts.  We plotted the weighted count
$\log[S^2 n(S)]$ instead of the traditional Euclidean-weighted count
$\log[S^{5/2}n(S)]$ as a function of $\log(S)$ because (1) $S^2 n(S)$
is proportional to the source contribution per decade of flux density
to the sky temperature (Eq.~\ref{dtbeq}), (2) the radio universe is
neither static nor Euclidean, and (3) the plotted slopes are minimized
for easy visual recognition of broad features, unlike the steeply
sloped plot of $\log[n(S)]$ for example.  On a plot of
$\log[S^2n(S)]$, the source count must ultimately fall off at both
ends to avoid Olbers' paradox.

The filled points at $\log[S\,{\rm (Jy)}] > -3$ are from the many
surveys referenced in \citet{con84a}, and the filled data points at
lower flux densities are from the \citet{mit85} confusion-limited VLA
survey. The irregular box encloses the range of 1.4~GHz counts
consistent with the probablility distribution $P(D)$ of confusion
amplitudes $D$ (in brightness units $\mu$Jy per beam solid angle) in
the low-resolution ($17\,\farcs5$ FWHM) \citet{mit85} image. The
straight line inside the box indicates the best power-law fit to the
$P(D)$ data.  The open data points and their power-law fit (upper
straight line) indicate the direct count of individual sources from
the most sensitive high-resolution 1.4 GHz survey ever made with the
original VLA \citep{owe08}.  These two faint-source counts disagree by
several times the Poisson counting errors, and nearly all other
published faint-source counts are scattered over the wide gap between
them \citep{con07}, as illustrated by Figure~11 in \citet{owe08}.  The
sources contributing to the data plotted in Figure~\ref{oldtbkndfig}
show no sign of resolving the extragalactic background at 1.4~GHz
because the power-law fits to $S^2 n(S)$ are still rising near
the flux density limit $S_0 \sim 10\,\mu$Jy.

The solid curve in Figure~\ref{oldtbkndfig} shows the \citet{con84b}
evolutionary model for the total source count at 1.4\,GHz. In this model,
the stronger radio sources are powered primarily by AGNs of
elliptical galaxies (dashed curve) and most of the fainter sources are in
star-forming spiral galaxies (dotted curve).  The model predicts that
(1) $S^2 n(S)$ should converge below $S \sim 10\,\mu$Jy and (2) most of
the model's $T_{\rm b} \approx 100$\,mK background should be
resolved into sources stronger than $S \sim 1\,\mu$Jy.  

However, \citet{fix11} reported that the recent ARCADE\,2 (Absolute
Radiometer for Cosmology, Astrophysics, and Diffuse Emission) balloon
measurement of absolute sky brightness leaves a remarkably high
extragalactic $T_{\rm b} = 54 \pm 6$\,mK at $\nu = 3.3$\,GHz after the
Galactic foreground and the $T = 2.731 \pm 0.004$\,K cosmic microwave
background (CMB) were subtracted.  Combining their 3.3~GHz result with
low-frequency data from the literature led \citet{fix11} to fit the
``excess'' extragalactic background with a power-law spectrum
\begin{equation}\label{arcade2tbeq}
T_{\rm b} = (24.1 \pm 2.1 {\rm ~K}) \times \biggl( {\nu \over
  310{\rm ~MHz}}\biggr)^{-2.599 \pm 0.036} 
\end{equation}
between 22\,MHz and 10\,GHz. Equation~\ref{arcade2tbeq} gives $T_{\rm
  b} \approx 480$\,mK at 1.4\,GHz, almost five times the $T_{\rm b}
\approx 100$\,mK predicted for the known populations of
extragalactic sources.

With the goals of (1) determining an accurate source count down to $S
\sim 1 \,\mu$Jy, (2) resolving the radio sky background contributed by
galaxies, and (3) constraining possible source populations that could
produce the high ARCADE\,2 background at 3.3\,GHz, we used the Karl
G.~Jansky Very Large Array (VLA) to make a very sensitive (rms noise
$\sigma_{\rm n} \approx 1\,\mu$Jy\,beam$^{-1}$) low-resolution ($8''$
FWHM) confusion-limited sky image covering one primary beam area at S
band (2--4\,GHz).  We pointed at the center of the ``crowded''
\citet{owe08} field to see if we could confirm the reported high
source count there.  Our relatively low angular resolution ensures
that galaxy-size sources at cosmological distances are unresolved:
$8''$ spans at least 40 kpc throughout the redshift range $0.4 < z <
7$ containing most faint radio sources.  That low resolution also
guarantees that the rms confusion is at least comparable with the rms
noise, a necessary condition for using confusion to constrain the sky
density of sources as faint as $S_0 \sim 1\,\mu$Jy, which is more than
a factor of five below our detection limit for individual sources.

\subsection{Outline}

Section~\ref{obssec} of this paper describes our observations and the
production of a confusion-limited sky image.  Section~\ref{pofdsec}
presents the 3.02~GHz $P(D)$ distribution from the
central region of this image.  The 1.4 and 3.02 GHz source counts
consistent with our $P(D)$ distribution are derived in
Section~\ref{countsec}, and the contributions of known extragalactic
source populations to the sky brightness are discussed in Section~\ref{tbsec}.
In Section~\ref{arcadesec} we use our narrow $P(D)$
distribution to show that the high ARCADE\,2 background is too smooth
to be produced by, or even spatially associated with, galaxies
brighter than $m_{\rm AB} = +29$. Section~\ref{summarysec} summarizes
our results.

\section{OBSERVATIONS AND IMAGING}\label{obssec}

\subsection{Target Field}

We reobserved the ``crowded'' \citet{owe08} field centered on J2000
$\alpha = 10^{\rm h}\,46^{\rm m}\,00^{\rm s}$, $\delta =
+59\arcdeg\,01\arcmin\,00''$ over the 2--4 GHz S-band frequency range
using an average of 21 antennas in the C configuration (maximum
baseline length $\approx 3$\,km) of the VLA.  This field in the
Lockman Hole was originally selected by \citet{owe08} because it was
covered by the {\it Spitzer} Wide-area InfraRed Extragalactic (SWIRE)
legacy survey \citep{lon03}, has exceptionally low infrared cirrus,
and contains no strong radio sources.  We favored it because it is far
enough north to minimize S-band radio-frequency interference (RFI)
from powerful geosynchronous broadcast satellites just south of the
celestial equator and has the highest reported faint-source count
\citep{owe08}.

\subsection{Observations}

Our 57 hours of observing time was divided among six observing nights
between 2012 February 21 and March 18, and 50 hours of this time was
spent integrating on the target field.  The nearby unresolved phase
calibrator J1035+564 was monitored for 30 seconds every 30 minutes.
The flux density and bandpass calibrators 3C 147 and 3C 286 were each
observed once per night.  Antenna pointing accuracy was maintained
with corrections determined from J1035+564 every hour.  Accurate
pointing is necessary even at S band because pointing errors can easily limit
the dynamic range of an image filled with sources \citep{con09}.

The data-averaging period was 1 second of time, and the 2--4~GHz
frequency range was divided into 16 correlator subbands, each with 64
spectral channels of width 2~MHz.  The spectral resolution was
broadened and the averaging intervals were increased somewhat prior to
imaging, but both bandwidth and time smearing were always
kept small enough that our sky coverage is limited only by the
primary-beam attenuation of the individual 25\,m antennas.  Their
attenuation pattern is very nearly that of a uniformly illuminated
circular aperture \citep{con98}:
\begin{equation} 
A(\rho /\theta_{\rm p}) \approx \biggl[{2 J_1 (3.233 \rho / \theta) 
\over (3.233 \rho / \theta)}\biggr]^2~,
\end{equation}
where $J_1$ is the Bessel function of the first kind and order, $\rho$
is the angular offset from the pointing center, and
\begin{equation}
\theta_{\rm p} = (43\farcm3 \pm 0\farcm4)
\,\biggl( {\nu \over {\rm GHz} }\biggr)^{-1}
\end{equation}
is the measured full width between half-maximum points (FWHM) of the
S-band primary beam at frequency $\nu$.  The primary FWHM ranges from
$21\farcm7$ at $\nu = 2$\,GHz to $10\,\farcm8$ at $\nu = 4$\,GHz.

\subsection{Editing and Calibration}

We used the Obit package
\citep{Obit}\footnote{http://www.cv.nrao.edu/$\sim$bcotton/Obit.html}
to edit and calibrate the $(u,v)$ data.  The six observing sessions
were calibrated and edited separately.  Two of the 16 subbands
contained satellite RFI strong enough to cause serious Gibbs ringing
in the raw data, a consequence of the finite number of correlator lags.
Hanning smoothing (combining adjacent spectral channels with weights
1/4, 1/2, and 1/4) suppressed this ringing and broadened the spectral
resolution to about 4 MHz. The new VLA correlator was designed to
prevent ``bleeding'' of ringing and other nonlinearities from one
subband to another.  Spectral channels still containing strong
interfering signals were flagged and removed from the data, as were a
few edge channels in each subband.  Prior to calibration, the
remaining data samples containing strong interfering signals that were
impulsive in time or frequency were identified by their large
deviations from running medians in time and frequency, and they were flagged.
Calibrator observations were also subjected to a test of the RMS/mean
amplitude ratios in each data stream, and noisy data segments with
high RMS/mean amplitude ratios were deleted.

Subsequent calibration and editing consisted of the following steps:
\begin{enumerate}
\item Instrumental group delay offsets were determined from
observations of 3C\,147, 3C\,286, and J1035+564 and applied to all of the data.
\item Residual variations of gain and phase with frequency were
corrected by bandpass calibration based on 3C\,286.
\item Amplitude calibration was based on the VLA standard spectrum of
  3C\,286 (see Table~1, column 3) bootstrapped to determine the
  spectrum of the astrometric calibrator J1035+564.  The more frequent
  observations of J1035+564 were then used to calibrate the amplitudes
  and phases the target $(u,v)$ data.  Data from some antennas, time
  intervals, and frequency ranges were still degraded by interference
  that had evaded earlier editing and corrupted a small fraction of
  our amplitude and phase solutions.  These corrupted solutions were
  detected by a comparison with all solutions, and data implying deviant
  complex gains were flagged.
\item The calibrated data were then subjected to further editing in which
  data with excessive Stokes I or V amplitudes were deleted, as well
  as another pass at removing narrowband interference.
\end{enumerate}

After this calibration and editing, the initial calibration was reset
and the whole process was repeated using only the data that had
survived the editing process.  Finally, because the target field contains no
sources brighter than 5\,mJy~beam$^{-1}$, we were able to flag the
small amount of the data having amplitudes significantly above the
noise.  About 53\% of the $(u,v)$ data survived all of the editing steps.
The calibrated and edited $(u,v)$ data from all six observing
sessions were then combined for imaging.  

\subsection{Imaging}

Observations spanning the frequency range between the low frequency
limit $\nu_{\rm l}$ and the high frequency limit $\nu_{\rm h}$ have 
center frequency $\nu_{\rm c} = (\nu_{\rm l} + \nu_{\rm h})/2$, 
bandwidth $\Delta \nu = (\nu_{\rm h} - \nu_{\rm l})$, and  fractional
bandwidth $\Delta\nu / \nu_{\rm c}$.  The fractional bandwidth covered
by our 2--4~GHz $(u,v)$ data is exceptionally large: $\Delta\nu /
\nu_{\rm c} = 2/3$. Using such data to make an image suitable for
measuring confusion encounters three interesting problems---the field
of view, the point-spread function (PSF), and the flux densities of
most sources can vary significantly with frequency.  To deal with
these problems, we separated the $(u,v)$ data into 16 subbands each having
a small fractional bandwidth $\Delta \nu / \nu_{\rm c} \ll 1$. The
$(u,v)$ data were tapered heavily in the higher-frequency subbands, and
each subband was imaged with an independent ``robustness''
\citep{bri95} to force nearly identical PSFs in all subbands.  We
weighted and recombined the narrowband images to produce a sensitive
wideband image characterized by an ``effective frequency'' $\langle
\nu \rangle$, where $\langle \nu \rangle$ at any position in a
wideband image is defined as the frequency at which the flux density
of a point source with a typical spectral index $\langle\alpha\rangle
\equiv d \ln S / d \ln \nu = -0.7$ equals its flux density in the
wideband sky image.  The effective frequency declines with angular
distance $\rho$ from the pointing center because the primary beamwidth
is inversely proportional to frequency.

 Table~1 lists the center frequencies $\nu_{\rm c}$ of the 16 subbands
 and the rms noise values $\sigma_{\rm n}$ of the 16 subband images.
 The fractional bandwidths of these subbands range from 3\% to 6\%, so
 each subband image is a narrowband image.  We used the Obit task
 MFImage to form separate dirty and residual images in each subband.
 We assigned each subband image a weight inversely proportional to its
 rms noise, generated a combined wideband
 image from their weighted average, and used this sensitive combined
 image to locate CLEAN components for a joint deconvolution.  The
 CLEAN operation used flux densities from the individual subband
 images but at locations selected from the combined image.  At the end
 of each major CLEAN cycle, the CLEAN components with flux densities
 from the individual subband images were used to create residual
 subband $(u,v)$ data and then new residual images.

The effects of sky curvature across the field of view were eliminated
by faceted ``fly's eye'' imaging in which each facet was small enough
to make the curvature effects negligible.  Each facet image was
projected onto a common coordinate grid on one tangent plane to
form a single continuous central image 30 arcmin in radius.  This
allowed all facets to be CLEANed in parallel.  Outside the central
image, separate outlier facets were added at the positions of those
NVSS \citep{con98} sources whose uncleaned sidelobes might otherwise
affect our target field.  Before imaging, the data were averaged
over baseline-dependent time intervals subject to the constraints that
(1) the averaging should not cause time smearing within our central
image and (2) the averaging time should never exceed the 20 second
phase self-calibration interval.

In order to obtain nearly identical PSFs in all subbands, we tapered
the $(u,v)$ data in each subband differently and assigned individual
``robust'' weighting factors adjusted to ensure that each synthesized
``dirty'' beam was nearly circular with major and minor axes between
$7''$ and $8''$.  After CLEANing, each residual image was smoothed by
convolution with its own elliptical Gaussian tailored to yield a
circular and nearly Gaussian PSF with a precisely $8\,\farcs0$ FWHM.  Having
the same dirty beam and PSF in each subband image is critical for an
accurate confusion analysis because only those sources much stronger
than the rms confusion were actually CLEANed.  Finally, all CLEAN
components were restored with an $8''$ FWHM circular Gaussian beam.

Two iterations of phase-only self calibration were used to remove
residual atmospheric phase fluctuations.  The final CLEAN cycle was
stopped when the peak in the combined residual image reached $10\,
\mu{\rm Jy\,beam}^{-1}$, a level well above the rms noise and
confusion in the combined image.  The total CLEANed flux density in
the combined image is 21 mJy.

Unwanted fluctuations indistinguishable from confusion are produced by
any dirty-beam sidelobes remaining in our combined image.  Fortunately, they are
small because the combination of long observing tracks and bandwidth
synthesis over our wide fractional bandwidth ensures excellent
$(u,v)$-plane coverage and keeps the dirty-beam sidelobe levels well
below 1\% of the peak, as shown in Figure~\ref{beamproflfig}.
Consequently the highest dirty-beam sidelobes from sources in the
residual image are $< 0.01 \times 10\,\mu{\rm Jy~beam}^{-1} \approx
0.1\,\mu{\rm Jy\,beam}^{-1}$, so their contribution to the image
variance is more than two orders of magnitude below the $\sim
(1\,\mu{\rm Jy~beam}^{-1})^2$ contributions of noise and
confusion.

We used the AIPS task IMEAN to calculate the rms noise values
$\sigma_{\rm n}$ of the CLEANed subband images in several large areas
that are well outside the main lobe of the primary beam and contain no
visible ($S_{\rm p} \gtrsim 6\,\mu{\rm Jy~beam}^{-1}$) sources; the
$\sigma_{\rm n}$ are listed in Table~1.  Next we assigned to each
subband image a weight inversely proportional to its noise variance
$\sigma_{\rm n}^2$, generated a wideband image from the weighted
average of the subband images, and measured the noise distribution in
four large regions well outside the primary main beam and free of
visible sources. Figure~\ref{noisehistfig} shows a logarithmic
histogram of the noise amplitudes in one such region covering an area
$\Omega = N \Omega_{\rm b}$ containing $N \approx 4000$ synthesized
beam solid angles $\Omega_{\rm b} \equiv \pi \theta_{\rm s}^2 /(4 \ln
2) \approx 0.020$\,arcmin$^2$ for $\theta_{\rm s} = 8''$.  The
excellent parabolic fit to the logarithmic histogram indicates a precisely
Gaussian noise amplitude distribution with an rms $\sigma_{\rm n}
\approx 1.02 \,\mu{\rm Jy\,beam}^{-1}$.

What is the rms statistical uncertainty $\Delta \sigma_{\rm n}$ in our
estimate of $\sigma_{\rm n}$?  The image PSF is a Gaussian, and the
effective noise area for a Gaussian PSF is that of the Gaussian PSF
squared \citep{con97,con98}.  Squaring a Gaussian PSF of width
$\theta$ yields a narrower Gaussian of width $2^{-1/2}\theta$ and
solid angle $\Omega_{\rm b}/2$.  Consequently there are actually $2N$
statistically independent noise samples in $\Omega = N \Omega_{\rm b}$
beam solid angles, and the rms fractional uncertainty in $\sigma_{\rm
  n}$ is
\begin{equation}\label{noisefluctuationeq}
{\Delta \sigma_{\rm n} \over \sigma_{\rm n}} = 
\biggl({1 \over 2N}\biggr)^{1/2}~, 
\end{equation}
not the commonly beliefed $(1/N)^{1/2}$. See the Appendix for a
detailed derivation of equation~\ref{noisefluctuationeq}.

The final result from the four areas covering a total $N = 11570$ beam
solid angles is $\sigma_{\rm n} = 1.012 \pm 0.007\, \mu{\rm
  Jy\,beam}^{-1}$ (statistical error only).  In theory the rms noise
is uniform across the image prior to correction for primary-beam
attenuation, so we used this value to estimate the noise in
confusion-limited regions near the pointing center.  The total
intensity-proportional error arising from uncertainties in the
flux-density calibration and primary beamwidth is not more than 3\%
inside the primary beam half-power circle.

\subsection{The SNR-Optimized Wideband Sky Image}

Our final wideband image of the sky was made with weights designed to
correct for primary-beam attenuation and simultaneously maximize the
signal-to-noise ratio (SNR) for sources having spectral indices near
$\langle\alpha\rangle = -0.7$, the mean spectral index of faint
sources found at frequencies around 3 GHz \citep{con84b}.  This
differs from the traditional weighting designed to minimize noise, which
maximizes the SNR in a narrowband image, but in a wideband image only
if $\langle\alpha\rangle \approx 0$.  The brightness $b_i (\rho)$ of
each pixel in each of the $i = 1, 16$ subband images was assigned a
weight
\begin{equation}\label{skyweighteq}
W_{\rm i}(\rho, \nu_{\rm c}) \propto \biggl[{\nu_{\rm c}^{\langle\alpha\rangle}
 \over \sigma_{\rm n} A(\rho,\nu_{\rm c})}\biggr]^2~,
\end{equation}
where the rms noise $\sigma_{\rm n}$ and center frequency $\nu_{\rm c}$
of each subband image is listed in Table~1.  
Each pixel in the weighted wideband sky image was generated from the ratio
\begin{equation}
b(\rho) = \sum_{i = 1}^{16}\, [b_i (\rho) W_i
  (\rho)]\,\bigg/\, \sum_{i=1}^{16} W_i(\rho)
\end{equation}
Even at the pointing center, weighting to optimize the SNR for
$\langle\alpha\rangle = -0.7$ increases the sky image noise slightly
from $\sigma_{\rm n} = 1.012 \pm 0.007\,\mu{\rm Jy~beam}^{-1}$ to
$\sigma_{\rm n} = 1.080 \pm 0.007\,\mu{\rm Jy~beam}^{-1}$.  Away from
the pointing center, the frequency-dependent primary-beam attenuation
correction causes the rms noise on the sky image to grow with the
radial offset $\rho$ at the rate indicated by the dashed curve in
Figure~\ref{skynoisefig}.  Weighting also affects how the effective
frequency $\langle \nu \rangle$ at each point in the wideband sky
image decreases with the offset $\rho$ from the pointing center.  The
dashed curve in Figure~\ref{skyfreqfig} indicates how $\langle \nu
\rangle$ in our sky image declines monotonically from 3.06 GHz at the
pointing center to 2.96 GHz at $\rho = 5$\,arcmin.

Figure~\ref{proflfig} is a profile plot of the central portion of our
SNR-optimized wideband sky image.  The intensity scale can be inferred
from the highest peaks, which are truncated at $S_{\rm p} =
100\,\mu{\rm Jy~beam}^{-1}$.  This image is confusion limited in the
sense that the rms fluctuations are everywhere larger than the noise
levels plotted in Figure~\ref{skynoisefig}.

%% In this section, we use  the \subsection command to set off
%% a subsection.  \footnote is used to insert a footnote to the text.


%% This section also includes several of the displayed math environments
%% mentioned in the Author Guide.

\section{THE $P(D)$ DISTRIBUTION}\label{pofdsec}

Historically, confusion was measured from the probability distribution
$P(D)$ of pen deflections $D$ on a chart-recorder plot of fringe
amplitudes from the single baseline of a two-element radio
interferometer \citep{sch57}.  On a single-dish or aperture-synthesis
array image, the corresponding ``deflection'' $D$ at any pixel is the
intensity in units of flux density per beam solid angle.  The observed
$D$ at any point is the sum of the contribution from the noise-free
source confusion and the contribution from image noise.  The source
confusion and noise contributions are independent of each other, so
the the observed $P(D)$ distribution is the convolution of the source
confusion and noise distributions, and the variance $\sigma_{\rm o}^2$
of the observed $P(D)$ distribution is the sum of variances of the
noise-free source confusion ($\sigma_{\rm c}^2$) and the noise
($\sigma_{\rm n}^2$) distributions:
\begin{equation}\label{sigmaoeq}
\sigma_{\rm o}^2 = \sigma_{\rm c}^2 + \sigma_{\rm n}^2~.
\end{equation}
The goal is to extract the source confusion distribution and its width
\begin{equation}\label{srcconfeq}
\sigma_{\rm c} = (\sigma_{\rm o}^2 - \sigma_{\rm n}^2)^{1/2}
\end{equation}
from the observed deflections and the measured noise.  If $\sigma_{\rm
  c} \ll \sigma_{\rm n}$, then $\partial \sigma_{\rm c} / \partial
\sigma_{\rm o} \approx (\sigma_{\rm o} / \sigma_{\rm c}) \approx
(\sigma_{\rm n} / \sigma_{\rm c}) \gg 1$ and small errors in the
measured $\sigma_{\rm o}$ (or $\sigma_{\rm n}$) cause large errors in
the extracted $\sigma_{\rm c}$.  This is the reason that only fairly
low-resolution $(\sigma_{\rm c} > \sigma_{\rm n}$) images can be used
to measure confusion.  Also, to the extent that other sources of error
(e.g., uncleaned sidelobes) are present but not accounted for,
equation~\ref{srcconfeq} will tend to overestimate the source
confusion.

The noise in our weighted sky image is low at the center but increases
with radial distance $\rho$ from the pointing center, and the noise
eventually overwhelms the confusion at large $\rho$.  On the other hand,
increasing the
radius $\rho$ of the circular region inside which the $P(D)$ distribution
is measured increases the number $N$ of beam solid angles sampled and thus
acts to decrease the statistical uncertainty $\Delta \sigma_{\rm o}$ in the
width of the observed $P(D)$ distribution.  For each thin ring of radius
$\rho$ covering $N$ beam solid angles, we use equation~\ref{sigmaoeq}
to calculate
%% see notes 20120801 in red/black notebook
%% begin revision: detailed calculation of (formerly numbered) equation 11
%% requested by referee's ``nitpick b''
\begin{equation}
(\Delta \sigma_{\rm o})^2 = \biggl( {\partial \sigma_{\rm o} \over 
\partial \sigma_{\rm c}} \Delta \sigma_{\rm c}\biggr)^2 +
\biggl( {\partial \sigma_{\rm o} \over 
\partial \sigma_{\rm n}} \Delta \sigma_{\rm n}\biggr)^2~,
\end{equation}
where
\begin{equation}
{\partial \sigma_{\rm o} \over \partial \sigma_{\rm c}} =
{\sigma_{\rm c} \over \sigma_{\rm o}} 
{\rm ~~~and~~~}
{\partial \sigma_{\rm o} \over \partial \sigma_{\rm n}} =
{\sigma_{\rm n} \over \sigma_{\rm o}}~.
\end{equation}
Then
\begin{equation}
\biggl( {\Delta \sigma_{\rm o} \over \sigma_{\rm o}} \biggr)^2 =
{1 \over \sigma_{\rm o}^4} 
\biggl[ \sigma_{\rm c}^4 
\biggl( {\Delta \sigma_{\rm c} \over \sigma_{\rm c}}\biggr)^2 +
\sigma_{\rm n}^4 
\biggl( {\Delta \sigma_{\rm n} \over \sigma_{\rm n}}\biggr)^2 \biggr]~,
\end{equation}
where 
$(\Delta \sigma_{\rm c} / \sigma_{\rm c}) \sim N^{-1/2}$
and 
$(\Delta \sigma_{\rm n} / \sigma_{\rm n}) = (2N)^{-1/2}$
(Eq.~\ref{noisefluctuationeq}).
In the limit $\sigma_{\rm n} \gg \sigma_{\rm c}$,
\begin{equation}
\biggl( {\Delta \sigma_{\rm o} \over \sigma_{\rm o}} \biggr)^2 \approx
\biggl( {1 \over 2N \sigma_{\rm o}^4 } \biggr) 
\sigma_{\rm n}^4~.
\end{equation}
%% end revision for ``nitpick b''
To minimize the uncertainty $\sigma_{\rm o}$ inside the circle of
radius $\rho$, we counted each pixel $\sigma_{\rm n}^{-4}$ times and
truncated the count at $\rho \approx 5$~arcmin, where $\Delta
\sigma_{\rm o} / \sigma_{\rm o}$ in the enclosed circle has a broad
minimum.  Within the circle of radius $\rho = 5$\,arcmin, our weighted
$P(D)$ distribution has rms noise $\sigma_{\rm n} = 1.255\,\mu{\rm
  Jy~beam}^{-1}$ (Fig.~\ref{skynoisefig}) and its effective frequency
is $\langle\nu\rangle \approx 3.02$\,GHz (Fig.~\ref{skyfreqfig}).

The weighted $P(D)$ distribution within 5\,arcmin of the pointing
center is shown by the data points with their Poisson rms error bars
in Figure~\ref{pofdbestfitfig}.  The thin curve indicates the Gaussian
noise distribution with rms width $\sigma_{\rm n} = 1.255\,\mu{\rm
  Jy~beam}^{-1}$.  Interferometers have no DC response to smooth
emission, so the mean deflection $\langle D \rangle \approx 3\,\mu{\rm
  Jy~beam}^{-1}$ in our image approximately equals the total CLEANed
flux density divided by the number of beam solid angles in the image.
Consequently, the image zero-point $D = 0$ should be treated as a free
parameter when comparing the observed $P(D)$ distribution with
analytic models of the true sky $P(D)$ distribution.

The noise-free confusion $P(D)$ distribution can be calculated
analytically for the scale-free case of a power-law differential
source count \citep{con74}
\begin{equation}
n(S) = k S^{-\gamma}~,
\end{equation}
where $k$ is the count normalization and $1 < \gamma < 3$ is the
differential count slope. The shape of the noise-free $P(D)$
distribution depends only on $\gamma$, and the width of the noise-free
$P(D)$ distribution obeys the scaling relation
\begin{equation}\label{pofdscalingeq}
P[(k \Omega_{\rm e})^{1/(\gamma-1)} D] = (k \Omega_{\rm e})^{-1/(\gamma-1)} P(D)~,
\end{equation}
where $\Omega_{\rm e}$ is the effective beam solid angle.  If $G$ is the
normalized ($G = 1$ at the peak) PSF, then
\begin{equation}\label{omegaeeq}
\Omega_{\rm e} = \int G^{\gamma -1} d \Omega
\end{equation} 
and
\begin{equation}
\Omega_{\rm e} = {\Omega_{\rm b} \over 
\gamma  -1}
\end{equation}
for a circular Gaussian PSF.  Equation~\ref{omegaeeq} also implies
that sidelobes of the PSF have a bigger effect as $\gamma$ declines,
so images suitable for confusion studies at sub-$\mu$Jy levels, where
$\gamma -1 \sim 0.5$, need to have very low sidelobe levels.  
The rms confusion $\sigma_{\rm c}$
contributed by deflections fainter than some limiting signal-to-noise
ratio $q = S_0 / \sigma_{\rm c}$ is \citep{con74}
\begin{equation}\label{rmsconfusioneq}
\sigma = \biggl({q^{3-\gamma} \over 3 - \gamma}\biggr)^{1/(\gamma-1)}
(k \Omega_{\rm e})^{1/(\gamma - 1)}~.
\end{equation}
If the number of sources per steradian stronger than $S$ is
\begin{equation}
N(>S_0) = \int_{S_0}^\infty n(S) dS
\end{equation}
then there are 
\begin{equation}\label{betadefeq}
\beta \equiv [N(>S_0)\Omega_{\rm b}]^{-1}
\end{equation}
beam solid angles per source stronger than $S_0$. For a power-law
source count,
\begin{equation}\label{beamspersourceeq}
\beta = {q^2 \over 3-\gamma}~.
\end{equation}
Taking $q = 5$ for reliable source detection \citep{mur73} and $\gamma
\approx 2$, equation~\ref{rmsconfusioneq} limits the maximum density
of reliably detectable and individually countable sources to about one
source per 25 beam areas if $\gamma \sim 2$.

These equations apply to unresolved sources.  
%% begin revision: Referee's ``nitpick b'' said the derivation of 
%% the equation below (formerly numbered 19) was unclear.  The
%% revised exlanation is: 
The response to an extended source is the
convolution of the source brightness distribution with the
point-source response, and areas add under convolution.  The
integrated flux density is conserved, so the peak flux density
falls. In the limit of a slightly resolved source covering a solid
angle $\Omega_{\rm s} \ll \Omega_{\rm e}$, the peak flux density will
be multiplied by the factor
%% end revision
\begin{equation}
f \approx {\Omega_{\rm e} \over \Omega_{\rm e} + \Omega_{\rm s}} \lesssim 1
\end{equation}
and the solid angle of the source response on the image will be divided by
$f$. Thus equation~\ref{pofdscalingeq} can be used to show that
slightly extended sources multiply the width of the $P(D)$
distribution by
\begin{equation}\label{widthscaleeq}
f^{(\gamma - 2)/(\gamma - 1)}~.
\end{equation}
%% see red book notes 2012 Jun 28

The thick curve in Figure~\ref{pofdbestfitfig} is the convolution of
the $\sigma_{\rm n} = 1.255\,\mu{\rm Jy~beam}^{-1}$ noise Gaussian
with the noise-free $P(D)$ distribution for the power-law source count
$n(S) = 9000 S^{-1.7}{\rm ~Jy}^{-1}{\rm ~sr}^{-1}$ of point sources
and an $8''$ FWHM Gaussian PSF. Even if faint sources are as large as
$\langle\theta_{\rm s}\rangle \approx 1\,\farcs5$ \citep{owe08},
equation~\ref{widthscaleeq} indicates that the width of the $P(D)$
distribution would increase by less than 2\%.  Thus the source count based on
our low-resolution $P(D)$ data should be much less sensitive to the
angular-size distribution of faint sources than direct source counts
from images that must have synthesized beam solid angles about 25
times smaller to reach the same flux-density limit.

\section{SOURCE COUNTS AT 1.40 AND 3.02 GHz} \label{countsec}

To compare our 3.02~GHz $P(D)$ distribution with source counts at
1.4~GHz, we assume that the average spectral index between 1.4 and
3.02~GHz is $\langle \alpha \rangle = -0.7$, the value that best fits
multifrequency counts of stronger sources \citep{con84b}. The
power-law count $k_1 S^{-\gamma}$ at $\nu_1 = 1.4$~GHz can be
converted to the count $k_2 S^{-\gamma}$ at $\nu_2 = 3.02$~GHz via the
relation
\begin{equation}
\biggl({k_1 \over k_2} \biggr) = 
\biggl({\nu_1 \over \nu_2}\biggr)^{\langle\alpha\rangle (\gamma-1)}
\end{equation}
If our estimate of $\langle \alpha \rangle$ is in error by 0.05 and
$\gamma = 1.7$, the frequency conversion will cause an $\approx 3$\%
error in the calculated value of $k_2$.

To span the range of faint-source counts reported at 1.4~GHz, we
converted the relatively low \citet{mit85} count $n(S) =
57\,S^{-2.2}{\rm ~Jy}^{-1}{\rm ~sr}^{-1}$ and the relatively high
\citet{owe08} count $n(S) = 6 S^{-2.5}{\rm ~Jy}^{-1}{\rm ~sr}^{-1}$ at
$\nu_1 = 1.4$~GHz to counts at $\nu_2 = 3.02$~GHz and extrapolated
them to lower flux densities.  The noise-free 3.02~GHz $P(D)$
distributions for these counts observed with a $\theta = 8''$ FWHM
Gaussian beam were convolved with the $\sigma_{\rm n} = 1.255\,\mu{\rm
  Jy~beam}^{-1}$ Gaussian noise distribution for comparison with the
observed $P(D)$ distribution (Figure~\ref{pofdsextrapfig}).  The continuous
curve is slightly lower and broader than the $P(D)$ data, indicating
that the extrapolated \citet{mit85} count is somewhat high, as
expected from the rapidly converging weighted count $S^2 n(S)$ of the
\citet{con84b} model.  The dashed curve is much lower and broader
because the extrapolated \citet{owe08} count is much higher.

Next we compared our 3.02 GHz $P(D)$ distribution with the predictions
of two 1.4\,GHz evolution models: 

(1) The very simple \citet{con84b} model is based on old local 1.4~GHz
luminosity functions derived from optically selected samples of
elliptical galaxies \citep{aur77} and spiral galaxies \citep{hum80}.
The radio sources in elliptical and spiral galaxies are primarily
associated with AGNs and star formation, respectively, so these
luminosity functions are similar to more modern luminosity functions
in which the FIR/radio correlation and/or optical line spectra are
used to distinguish radio sources primarily powered by AGN from those
primarily powered by recently formed stars \citep{con02,mau07}.  There
were only sparse redshift data for identifications of strong radio
sources, and a pre-WMAP cosmology was assumed.  The decision to evolve
both the AGN and star-forming source populations identically to match
the source count at several frequencies was based on applying Occam's
razor to limited source-count data below $S \sim 1$~mJy, and not any
physical theory.

(2) The far more sophisticated and modern \citet{wil08} semi-empirical
simulation of the sky was made as part of the Square Kilometer Array
Design Study (SKADS).  It uses new and better local luminosity
functions, redshift data, and a concordance cosmological model
incorporating dark matter and dark energy.  

Remarkably, both models predict nearly identical source counts all the
way down to $S \sim 10$\,nJy, and both agree well with our observed
$P(D)$ distribution.  This good agreement suggests that there is
little ``wiggle room'' for models describing the counts of faint
sources once the local radio luminosity function has been specified;
the details of the evolution and the underlying cosmology seem to have
little effect on the resulting source count \citep{con89}.  The fact
that our $P(D)$ count matches so well the $S^2 n(S)$ convergence below
$S \approx 10\,\mu{\rm Jy}$ predicted by the \citet{con84b} and
\citet{wil08} models provides independent support for the claim that
most faint radio sources are in star-forming galaxies.  Such radio
sources are usually somewhat smaller than their optical host galaxies,
so we expect that the median angular diameter of $\mu$Jy sources is
$\theta_{\rm s} \leq 1''$.

In the flux-density range $1 \lesssim S \lesssim 10 \,\mu$Jy, the 1.4
GHz model counts can be approximated by the power laws $n(S) =
1.2\times10^5 S^{-1.5} {\rm ~Jy}^{-1} {\rm ~sr}^{-1}$ \citep{con84b}
and $n(S) = 3.45\times10^4 S^{-1.6} {\rm ~Jy}^{-1}{\rm ~sr}^{-1}$
\citep{wil08}.  At 3.02 GHz these counts become $n(S) = 9.17 \times
10^4 S^{-1.5} {\rm ~Jy}^{-1}{\rm ~sr}^{-1}$ and $n(S) = 2.5 \times
10^4 S^{-1.6} {\rm ~Jy}^{-1}{\rm ~sr}^{-1}$, respectively, if $\langle
\alpha \rangle = -0.7$. Figure~\ref{pofdsmodelsfig} compares the
\citet{con84b} (dashed curve) and \citet{wil08} (continuous curve)
model $P(D)$ distributions, after conversion to 3.02~GHz with $\langle
\alpha \rangle = -0.7$ and convolution with $\sigma_{\rm n} =
1.255\,\mu{\rm Jy~beam}^{-1}$ noise, with the 3.02~GHz $P(D)$ data.
  
Although the power-law approximations to both models appear to agree
fairly well with the $P(D)$ data in Figure~\ref{pofdsmodelsfig}, the
normalized $\chi^2_\nu$ values from fits between $D = -3.9\,\mu{\rm
  Jy~beam}^{-1}$ and $D = +9.7\,\mu{\rm Jy~beam}^{-1}$ ($\nu = 66$
degrees of freedom) are too high: $\chi^2_\nu \sim 100$ for the
power-law approximation to the \citet{con84b} model and $\sim 10$ for
the power-law approximation to the \citet{wil08} model.  Both
approximations significantly overestimate the numbers of sources
stronger than about $5\,\mu$Jy.  The $P(D)$ data are simply not
consistent with {\it any} power-law approximation to the source count
over the whole range $1 \lesssim D(\mu{\rm Jy~beam}^{-1}) \lesssim
10$.  The best-fit slope at the high end of this range is $\gamma
\approx 1.7$ while the best-fit slope at the low end is $\gamma
\approx 1.5$.  Better modeling of the source counts will require
numerical simulations to generate $P(D)$ distributions for
non-power-law source counts.  The range of 3.02 GHz power-law counts
reasonably consistent with the $P(D)$ data is spanned by the three
fits already mentioned: 
%% begin revision: exponent corrected from -4 to 4 (referee's ``nitpick c'')
$n(S) = 9.17 \times 10^4 S^{-1.5} {\rm
  ~Jy}^{-1}{\rm ~sr}^{-1}$
%% references added to these counts for clarity, following 
%% referee's ``nitpick c''
\citep{con84b}, $n(S) = 2.5 \times 10^4 S^{-1.6}{\rm
  ~Jy}^{-1}{\rm ~sr}^{-1}$ \citep{wil08}, and $n(S) = 9000 S^{-1.7}{\rm
  ~Jy}^{-1}{\rm ~sr}^{-1}$ (Fig.~\ref{pofdbestfitfig}).
%% end revision

\subsection{The Noise-Free RMS Confusion }

The corresponding noise-free $P(D)$ distributions for a $\theta = 8''$
FWHM Gaussian PSF are plotted in Figure~\ref{sourcepofd}.  Note that
the zero-point deflection $D = 0$ in these calculated $P(D)$
distributions is the absolute zero of sky brightness contributed by
discrete sources.  For values of $\gamma \leq 1.7$, all sources
fainter than $S \sim 1\,\mu$Jy contribute very little to the sky
background.  The rms confusion $\sigma_{\rm c}$ is a poor statistic
for describing such skewed $P(D)$ distributions.  In a Gaussian
distribution, about $2/3$ of the points lie within $\pm 1 \sigma$ of
the mean, so we use $\sigma_{\rm c}^*$ defined as half the range of
$D$ containing $2/3$ of the points as a more stable measure of the
width of the confusion distribution.  In a $\theta = 8''$ FWHM PSF,
this width of the noiseless confusion distribution is $ \sigma_{\rm
  c}^* \approx 1.2\,\mu{\rm Jy~beam}^{-1}$ at $\nu = 3.02$~GHz.  For
nearby values of $\theta$ and $\nu$, the scaling
%% begin revision: corrected wrong equation reference (referee's ``nitpick d'')
equation~\ref{rmsconfusioneq} 
%% end revision
with $\gamma \approx 1.6$ implies
\begin{equation}\label{confapproxeq}
\sigma_{\rm c}^* \approx 
1.2\,\mu{\rm Jy~beam}^{-1}
\biggl({ \nu \over {\rm 3.02~GHz}}\biggr)^{-0.7}
\biggl({ \theta \over 8''}\biggr)^{10/3}~.
\end{equation}
For example, the proposed EMU (Extragalactic Map of the Universe)
survey will cover most of the sky at $\nu = 1.3$~GHz with $\theta =
10''$ resolution \citep{nor11} and rms noise $\sigma_{\rm n} =
10\,\mu{\rm Jy~beam}^{-1}$.  Equation~\ref{confapproxeq} indicates
that EMU will not be confusion limited, with $\sigma_{\rm c}^* \approx
5\,\mu{\rm Jy~beam}^{-1}$.  

\subsection{The Natural Confusion Limit}

In addition to the instrumental confusion limit caused by finite
angular resolution, there is a ``natural'' confusion limit caused by
the finite angular size of faint sources.  If the median solid angle
covered by a faint source is $\langle\Omega_{\rm s}\rangle$, then
equations~\ref{betadefeq} and \ref{beamspersourceeq} imply the number
of source solid angles per detectable source ($q > 5$) must exceed
\begin{equation}
\beta =  [N(>S)\Omega_{\rm s}] = {q^2 \over 3-\gamma} \sim 25
\end{equation}
even if the instrumental resolution is very high.  Their high source
count prompted \citet{owe08} to speculate that the natural confusion
limit could be as high as $1\,\mu$Jy, but our lower $P(D)$ count
suggests that the natural confusion limit for detecting individual
$\langle \Omega_{\rm s} \rangle \sim (1'')^2$ sources is under
10~nJy.  This is well below the hoped-for $5\sigma_{\rm n} = 45$\,nJy
sensitivity limit of the full 12000~m$^2$~K$^{-1}$ Square Kilometre
Array (SKA) after 1 megasecond of integration, so the SKA continuum
sensitivity should not be limited by either natural or instrumental
confusion, but only by instrumental noise and dynamic range \citep{con09}.

\subsection{The 1.4~GHz Source Count}

The full range of 1.4~GHz source count data is plotted with the
traditional $S^{5/2} n(S)$ ``static Euclidean'' normalization in
Figure~\ref{eucfig} for direct comparison with most published results.
The range of 1.4~GHz source counts consistent with our $P(D)$ distribution is
conservatively within the box covering $2 \lesssim S \lesssim
20\,\mu{\rm Jy}$.  Again, obtaining more precise constraints on the
counts consistent with our $P(D)$ distribution will require numerical
simulations of non-power-law counts.

The upper abscissa of Figure~\ref{eucfig} shows the source-frame
1.4~GHz spectral luminosity $L({\rm W~Hz}^{-1})$ of a normal-spectrum
($\alpha \approx -0.7$) source whose flux density is directly below on
the lower abscissa if the source is at the median redshift $\langle z
\rangle \sim 0.8$ of extragalactic sources \citep{con89}.  These
typical luminosities and flux densities at $z \sim 0.8$ can be
compared with those of nearby radio sources such as the Large
Magellanic Cloud (LMC) at $\log(L) \approx 20.3$ 
%% begin revision: added three references for the luminosities, as
%% requested by referee's ``nitpick e''
\citep{hug07} 
and $\log(S) \approx -7.1$, our own Galaxy at $\log(L) \approx 21.4$ 
\citep{ber84}
and $\log(S) \approx -6.0$, M82 at $\log(L) \approx 22.1$
\citep{kel69}
and $\log(S) \approx -5.3$, and Arp 220 at $\log(L) \approx 23.4$ 
\citep{con02}
%% end revision 
and $\log(S) \approx -4.0$.  Most of the extragalactic
radio sources fainter than our $P(D)$ limit were no more luminous at
$z \sim 0.8$ than our Galaxy is today.  For the first time, the count
data are deep enough to confirm the assumption \citep{con84a,con84b}
that radio sources powered by AGNs and radio sources powered by star
formation evolve at about the same rate.  That assumption had no known
physical justification in 1984, but it is consistent with the
correlation between black hole and stellar bulge masses \citep{mag98}
discovered later.  
%% begin revision, inserted to address the referee's ``main concern''
Note that this correlation appears to be weaker for the lower-mass
($\sim 10^7 ~M_\odot$) black holes found in late-type star-forming
galaxies \citep{gre10}.  The processes that produced the correlation
for massive elliptical galxies do not appear to have had enough time
to fully establish the correlation for the star-forming galaxies that
host most faint radio sources.
%% end revison

Our new source count is about a factor of four lower than the
\citet{owe08} count near $S = 15\,\mu$Jy, even though our fields
overlap on the sky.  What might cause this discrepancy and, for that
matter, the surprisingly large scatter among all published
faint-source counts \citep{con07}?  Are most of the faint
\citet{owe08} sources spurious, or did we miss a large fraction of
real sources that \citet{owe08} counted?  The \citet{owe08} image has
$\theta = 1\,\farcs6$ angular resolution, much higher than our $\theta
= 8''$ resolution and the \citet{mit85} $\theta = 17\,\farcs5$
resolution.  How often have we blended into one ``source'' what
\citet{owe08} resolved into two or more sources?  Figure~\ref{grayfig}
shows our 3~GHz sky image as a gray scale with white crosses on the
positions of cataloged \citet{owe08} sources stronger than 15$\,\mu$Jy
at 1.4~GHz.  The correspondence is actually quite good.  Almost all
crosses have 3~GHz counterparts, and most of the brighter 3~GHz
sources have 1.4~GHz counterparts, so spurious or missing sources
cannot explain a factor of four discrepancy.  Likewise, there are only
a few cases of two \citet{owe08} sources blending into one source on
our low-resolution image.  We suspect that most of the count
difference is caused by count corrections made for partial resolution of
extended sources in the high-resolution 1.4~GHz beam.  Survey catalogs
are complete to a fixed brightness ($\mu$Jy~beam$^{-1}$) cutoff, so
extended sources with lower brightnesses but higher integrated flux
densities ($\mu$Jy) will be missed.  The corrections needed to convert
source brightnesses in $\mu{\rm Jy~beam}^{-1}$ to source flux
densities in $\mu$Jy and to account for missing sources become quite
large near the brightness cutoff as the angular resolution approaches
the median angular size of faint sources.  For example, \citet{owe08}
found a median angular size $\langle \phi \rangle = 1\,\farcs5$ in
their 20--30~$\mu$Jy bin, which is about the same as their FWHM
resolution $\theta = 1\,\farcs6$. The rapid rise in counts near
cutoffs at 6, 7, 8, 9, and 10 times the rms noise shown in Figure~9 of
\citet{owe08} may be a sign that these corrections were too large.

\section{THE CONTRIBUTIONS OF KNOWN SOURCES TO THE SKY BRIGHTNESS}
\label{tbsec}

The 1.4 GHz source counts in Figure~\ref{eucfig} were replotted as
$\log[S^2 n(S)]$ versus $\log(S)$ in Figure~\ref{tbcountfig} to
emphasize the source contribution to sky brightness.  The right
ordinate of the top panel in Figure~\ref{tbcountfig} shows the
differential contribution $d T_{\rm b} / d[\log(S)]$ to the 1.4~GHz sky
brightness temperature per decade of flux density.  The constraint
provided by the new $P(D)$ distribution clearly shows that $S^2 n(S)$
is falling off at low flux densities, at about the rate predicted by the
\citet{con84b} and \citet{wil08} evolutionary models.

We estimated the 1.4~GHz and 3.02 GHz backgrounds from the known
source populations powered by AGNs and star-forming galaxies by
inserting the 1.4 GHz source counts indicated by the solid curve in
the top panel of Figure~\ref{tbcountfig} into equation~\ref{tbcumeq}
and assuming an effective spectral index $\langle \alpha \rangle =
-0.7$.  The results are shown in the middle and bottom panels,
respectively.  The total background from the model sources converges
to $T_{\rm b} \approx 100$\,mK at 1.4~GHz and $T_{\rm b} \approx
13$\,mK at 3.02~GHz.  AGNs account for about 63\% and star-forming
galaxies 37\% of the extragalactic source background.  Approximately
96\% of this total has been resolved into sources stronger than $S_0
\approx 2 \,\mu{\rm Jy}$ at 1.4~GHz or $S_0 \approx 1 \,\mu{\rm Jy}$
at 3.02~GHz.  More specifically, nearly 100\% of the AGN background
and about 89\% of the star-forming galaxy contribution to the
background has been resolved.

Star-forming galaxies appear to obey the local FIR/radio correlation
even at moderate redshifts $z \lesssim 2$ \citep{ivi10}, while
the radio sources powered by AGNs have much lower FIR/radio flux
ratios.  Local ($z \ll 1$) star-forming galaxies have a mean FIR/radio
flux ratio $\langle q \rangle \equiv \log[S(80\,\mu{\rm m}) / S({\rm
    1.4~GHz})] \approx 2.3$ \citep{con92}.  Our claim that
star-forming galaxies produce a $T_{\rm b} \approx 37\,{\rm mK}$
background at $\nu = 1.4$~GHz (equivalently, $I_\nu \sim 2.2 \times
10^{-23}{\rm ~W~m}^{-2}{\rm ~Hz}^{-1} {\rm ~sr}^{-1}$) can be checked
by comparison with the measured extragalactic FIR background.
\citet{fix98} and \citet{lag99} analyzed COBE FIRAS data to find $\nu
I_\nu = 13.7 \pm 4{\rm ~nW~m}^{-2}{\rm~sr}^{-1}$ at $\lambda =
160\,\mu$m, or $I_\nu \approx 7.3 \times 10^{-21}{\rm ~W~m}^{-2}{\rm
  ~Hz}^{-1}{\rm ~sr}^{-1}$ at $\nu \approx 1870$~GHz.  Thus the
background FIR/radio ratio is $q_{\rm b} \equiv
\log[I_\nu(160\,\mu{\rm m}) / I_\nu({\rm 1.4~GHz})] \approx 2.5$.  If the
  star-forming background galaxies have typical redshifts $\langle z
  \rangle \sim 0.8$ and radio spectral indices $\alpha \approx -0.7$,
  then these backgrounds imply that their source-frame FIR/radio ratio
  is $ \log[S(90\,\mu{\rm m}) / S(2.5{\rm ~GHz})] \approx 2.5$, which
  corresponds to a source-frame FIR/radio flux ratio
  $\log[S(90\,\mu{\rm m}) / S(1.4{\rm ~GHz})] \approx 2.3$.  This
  agrees with the local FIR/radio correlation and supports the idea
  that the same population of galaxies accounts for the star-forming
  radio and FIR backgrounds.

The Herschel PACS Evolutionary Probe (PEP) resolved $74\pm5$\,\% of
the $\lambda = 160\,\mu$m background into individually detected
galaxies stronger than about 3~mJy, and \citet{ber11} used a $P(D)$
analysis to raise the resolved fraction to $\sim 89$\% above $\sim
0.3$\,mJy.  The sensitivity limits of the VLA ($\theta = 8''$ FWHM
resolution) and Herschel PEP $\lambda = 160\,\mu$m ($\theta \approx
11''$ FWHM resolution) $P(D)$ distributions to galaxies obeying the
FIR/radio correlation are about equal, and the same fraction of the
star-forming galaxy background has been resolved at radio and FIR
wavelengths.  Converting the direct and $P(D)$ counts at $\lambda =
160\,\mu$m to 1.4~GHz using the relation $\log[S(160\,\mu{\rm m}) /
  S({\rm 1.4~GHz})] = 2.5$ allows them to be compared directly with
the \citet{con84b} 1.4~GHz count model for star-forming galaxies
(dotted curve in Fig.~\ref{bertacountfig}); the agreement is well
within the statistical errors.

\section{THE ARCADE\,2 EXTRAGALACTIC BACKGROUND and 
CONSTRAINTS ON NEW POPULATIONS OF RADIO SOURCES}\label{arcadesec}

If the extragalactic source background is really $T_{\rm b} \sim
100$\,mK at 1.4~GHz and 13~mK at 3.02~GHz, then it has been resolved
into the known source populations powered by AGNs and star formation.
However, all of the faint-source counts are based on interferometric
images which are insensitive to a smooth background, the 2.7~K CMB for
example.  We cannot rule out a sufficiently smooth distribution of
faint radio sources that contributes significantly to the
extragalactic background brightness, but we can set strong lower
limits to the numbers of sources needed to make the background smooth
enough to agree with our narrow 3.02~GHz $P(D)$ distribution.

The ARCADE\,2 measurement of the extragalactic sky temperature in the
3--90~GHz frequency range \citep{fix11} recently reported a surprising
extragalactic excess over the CMB. Its brightness spectrum is
\begin{equation}\label{arcade2eq}
\biggl({ T_{\rm b} \over {\rm K}}\biggr) = (24.1 \pm 2.1) \,
   \biggl({ \nu \over 0.31\,{\rm GHz}}\biggr)^{-2.599 \pm 0.036}
\end{equation}
between 22\,MHz to 10~GHz.  This is $T_b
= 480 \pm 50$\,mK at $\nu = 1.4$~GHz and $T_{\rm b} = 65 \pm 8$\,mK
at $\nu = 3.02$~GHz, much higher than the $100\pm 10$\,mK and $13 \pm 1.3$\,mK
attributable to known extragalactic radio sources at 1.4~GHz and 3.02~GHz,
respectively.  

One possible explanation is the existence of a new
population of faint extragalactic sources that contributes $\Delta T_{\rm b}
\approx 480 - 100 = 380 \pm 50$\,mK at 1.4~GHz and $\Delta T_{\rm b} \approx
65 - 13 = 52 \pm 8$\,mK at 3.02~GHz to the sky background.
A truly diffuse synchrotron background produced by
relativistic electrons in intergalactic space far from individual galaxies
was ruled out by \citet{sin10} on the grounds that inverse-Compton scattering
of those electrons radiating in weak ($B < 1\,\mu$G) intergalactic
magnetic fields would exceed the observed X-ray and $\gamma$-ray
backgrounds.  \citet{sin10} also pointed out that the diffuse emission from
clusters of galaxies has a spectrum that is too steep ($\alpha < -1$) to
explain the ARCADE\,2 excess, as does emission from shocks in the cosmic
web \citep{bro11}.

Equation~\ref{tbcumeq} constrains the source count of any new
population of faint unresolved sources contributing $\Delta T_{\rm b}$ to the
sky brightness.  Figure~\ref{bumpfig} plots $\log[S^2n(S)]$ as a
function of $\log(S)$ for the known populations of radio sources
powered by AGN and by star-forming galaxies; they appear as two broad
peaks centered near $\log(S_{\rm pk}) \sim -1$ and $\log(S_{\rm pk}) \sim
-5$, respectively.  The regions near these peaks
contribute most to the sky brightness, and each peak is well represented by
the parabolic Taylor-series approximation $\log[S^2 n(S)] \approx a -
b [\log(S) - \log(S_{\rm pk})]^2$ or
\begin{equation}\label{gausseq}
S^2 n(S) \approx A \exp \biggl\{- 4 \ln(2) {[\log(S)- \log(S_{\rm pk})]^2 
\over \phi^2} \biggr\}~,
\end{equation}
where $\phi$ is the logarithmic FWHM of the Gaussian and $S_{\rm pk}$
is the flux density of the peak.  The peak amplitude $A$
%% begin revision: deleted ``the logarithm of'' before ``the peak 
%% brightness (referee's ``nitpick f'')
(the peak brightness in units of Jy~sr$^{-1}$)
%% end revision
required for a new population characterized by a Gaussian of FWHM
$\phi$ to add $\Delta T_{\rm b}$ to the background can be found by inserting
equation~\ref{gausseq} into equation~\ref{tbcumeq} and integrating
over all flux densities ($S_0 = 0$); it is
\begin{equation}\label{newpopeq}
A\phi = {4 k_{\rm B} \nu^2  \over \ln(10) c^2} 
\biggl[{ \ln(2) \over \pi}\biggr]^{1/2} \Delta T_{\rm b}~.
\end{equation}
$A$ and $\phi$ are individually free, but their product $A \phi$ is fixed
by $\Delta T_{\rm b}$.  

Figure~\ref{bumpfig} shows three examples of source counts, each of
which yield $\Delta T_{\rm b} \approx 380$\,mK at $\nu = 1.4$~GHz.
The dotted, dashed, and continuous curves correspond to logarithmic
FWHMs $\phi = 0.2$,\,1.0, and 2.0 in equation~\ref{gausseq}, and their
peak amplitudes calculated from equation~\ref{newpopeq} are
$\log[A({\rm Jy\,sr}^{-1})] \approx 4.67$, 3.97, and 3.67,
respectively.  The value of $\Delta T_{\rm b}$ is independent of
$S_{\rm pk}$, so these parabolic curves are free to shift horizontally
when constrained by sky brightness alone.  However, a large value of
$S_{\rm pk}$ decreases the number of sources per square arcmin needed
to produce the excess background and increases the Poisson fluctuations
in sky brightness, so the width of our observed $P(D)$ distribution
sets upper limits to the values of $S_{\rm pk}$ associated with these
three examples.  If the background is too bright and smooth, the
number of faint radio sources required will exceed the total number of
galaxies.

The narrowest possible peak ($\phi \ll 1$) needs the smallest number
$N$ of sources per beam solid angle to produce a given $T_{\rm b}$.
In that unrealistic limit, all of the new sources have nearly the same
very low flux density $ S_{\rm pk}$ and their noiseless $P(D)$
distribution is nearly Gaussian.  To set a very conservative upper limit to
the rms width of this Gaussian, we found the widest Gaussian
consistent with our observed $P(D)$ distribution.  It is shown in
Figure~\ref{pofdsbump}, and its rms is $\sigma \approx 0.70\,\mu{\rm
  Jy~beam}^{-1} \approx 1.47$\,mK at $\nu = 3.02$~GHz.  To avoid
excessive Poisson noise, the minimum number $N$ of sources per beam
solid angle $\Omega_{\rm b} \approx 0.020$~arcmin$^2$ must be
\begin{equation}
N > \biggl({\Delta T_{\rm b} \over \sigma}\biggr)^2 =
\biggl({52{\rm ~mK} \over 1.47{\rm ~mK}}\biggr)^2 \approx 1.3\times10^3~,
\end{equation}
implying $>6\times10^4$ radio sources per arcmin$^2$, and the maximum
source flux density consistent with the background brightness is
$S_{\rm pk} < 34$\,nJy at 1.4~GHz if $\alpha = -0.7$.  This is
slightly below the hoped-for $5\sigma_{\rm n} = 45$\,nJy sensitivity
limit of the full 12000~m$^2$~K$^{-1}$ SKA after 1 megasecond of
integration, so even if such faint sources exist, it will be a long
time before they can be detected individually.

Broader flux distributions can only raise the minimum source density
and lower the peak flux density.  If $\phi = 1$ ($\phi=2$), then there
must be more than $1.6\times10^5$ ($3\times10^6$) sources per
arcmin$^2$ with peaks at $S_{\rm pk} < 22$\,nJy ($<5$\,nJy).  The dotted,
dashed, and continuous curves in Figure~\ref{bumpfig} are as far right
as allowed by the $P(D)$ smoothness constraint.  They could be shifted
to the left, but that would only increase the implied sky density of
radio sources.  Our new constraint on the smoothness of the background
forces the ``bump'' in the weighted source count at $\mu$Jy levels
proposed by \citet{sei11} and \citet{ver11} to much lower flux levels.

There are only $10^4$ galaxies brighter than $m_{\rm AB} = +29$ in the
11~arcmin$^2$ Hubble Ultra Deep Field (HUDF) \citep{bec06}, or
$\approx 10^3$ galaxies per arcmin$^2$, about two orders magnitude
lower than the minimum sky density of faint radio sources needed to
produce the smooth ARCADE\,2 excess background.  The HUDF can detect
galaxies 2~mag fainter than $L^*$ out to redshifts $z \approx 6$, so
it appears that the smooth ARCADE\,2 background cannot be produced by
galaxies or by objects located in or near individual galaxies (e.g.,
multiple radio supernovae or gamma-ray bursts in galaxies, radio-quiet
quasars, radio halos smaller than 8 arcsec in diameter) or by
star-forming galaxies that are more radio-loud than expected from the
FIR/radio correlation \citep{sin10}.  Might faint sources produced by
WIMP annihilations or decays in dark-matter (DM) halos \citep{for11}
be sufficiently numerous to produce a sufficiently smooth background?
Annihilation in DM halos around galaxies would violate our smoothness
constraint, leaving only the possibility of preferential DM
annihilation in high-redshift mini-halos.  If our 3.02~GHz $P(D)$
smoothness constraint and the \citet{fix11} 3.3~GHz excess brightness
are both correct, then a very numerous ($N > 10^{13}$ over the whole
sky) and unexpected population of radio sources not associated with
galaxies has been discovered.

\section{SUMMARY}\label{summarysec}

We used 21 antennas in the Karl G.~Jansky VLA to observe a single
field at S band (2--4~GHz) with a FWHM resolution $\theta =
8''$ and reached an rms noise $\sigma_{\rm n} \approx 1\,\mu{\rm
  Jy~beam}^{-1}$ near the image center after 50 hours of integration
time. The image is confusion limited with an ``rms'' confusion level
$\sigma_{\rm c}^* \approx 1.2\,\mu{\rm Jy~beam}^{-1}$ at $\nu =
3.02$~GHz.

The 3.02~GHz differential source count was derived from the confusion
$P(D)$ distribution.  For comparison with published source counts, we
converted it to 1.4~GHz via the effective spectral index $\langle
\alpha \rangle \approx -0.7$.  The power-law approximation $n(S)
\propto S^{-\gamma}$ to the 1.4~GHz source count has a slope
approaching $\gamma \approx 1.5$ near $1~\mu$Jy, significantly lower
than the slopes of published counts above $10~\mu$Jy.  If the faintest
radio sources have median angular size $\langle \theta_{\rm s} \rangle
\leq 1''$ as expected for emission coextensive with star-forming
regions in distant galaxies, the natural confusion limit for source
detection is not more than $5\sigma_{\rm c}^* \approx 0.01\,\mu$Jy at
1.4~GHz, and the continuum sensitivity of the planned SKA will not be
limited by natural confusion.  The observed count is well fit by
evolutionary models in which the local radio luminosity functions of
all sources associated with both AGNs and star formation evolve at the
same rate.  This is broadly consistent with the correlation of black
hole and stellar bulge masses in massive elliptical galaxies, 
%% begin revision inserted to address the referee's ``main concern''
although this correlation is not yet well established for the
lower-mass black holes and bulges in late-type galaxies \citep{gre10}.
%% end revision

The brightness-weighted count $S^2 n(S)$ is clearly converging below
$10\,\mu$Jy.  Our image has resolved about 96\% of the radio
background produced by all galaxies ($T_{\rm b} \approx 100$~mK at
1.4~GHz and $T_{\rm b} \approx 13$~mK at 3~GHz).  Nearly 100\% of the
$\approx 63$~mK AGN-powered background at 1.4~GHz has been resolved.
The remaining $\approx 37$~mK comes from star-forming galaxies that
obey the FIR/radio correlation and account for most of the
extragalactic background at $\lambda = 160\,\mu$m. We resolved about
89\% of the star-forming galaxy contribution.

The ARCADE\,2 balloon experiment indicated a nonthermal excess
brightness over the Galaxy, the CMB, and that expected from known
populations of radio sources in galaxies.  At 3.02 GHz this excess
brightness temperature is $52 \pm 8$~mK. Our narrow 3.02~GHz $P(D)$
distribution implies that the excess background must be very smooth.
Any new discrete-source population able to produce such a bright and
smooth background is far too numerous to be associated with galaxies
brighter than $m_{\rm AB} = +29$.

%% The equation environment wil produce a numbered display equation.


%% Included in this acknowledgments section are examples of the
%% AASTeX hypertext markup commands. Use \url without the optional [HREF]
%% argument when you want to print the url directly in the text. Otherwise,
%% use either \url or \anchor, with the HREF as the first argument and the
%% text to be printed in the second.

%% \acknowledgments


%% in parentheses, after the keyword, but they will not be verified.

{\it Facilities:} \facility{VLA}



%% begin revision addressing referee's ``second concern''
\appendix

\section{Noise Distributions in Synthesis Images}

\begin{figure}
\vskip .5in
\includegraphics[angle=-90, scale= 0.7]{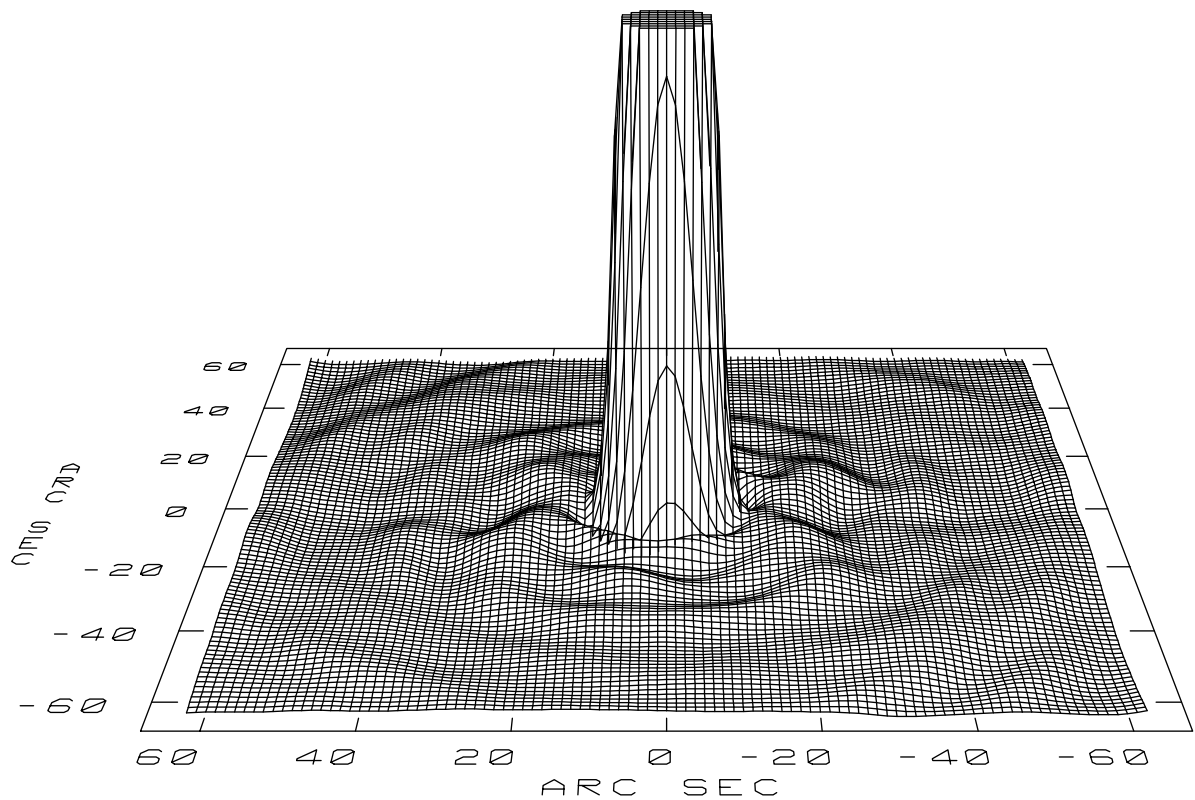}
\caption{Dirty-beam profile.  The dirty beam in our wideband image, shown
  truncated at 20\% of its peak to demonstrate that the highest
  sidelobe level is under 1\%.}
\label{beamproflfig}
\end{figure}

\begin{figure}
\vskip .5in
\includegraphics[angle=-0, scale= 0.7]{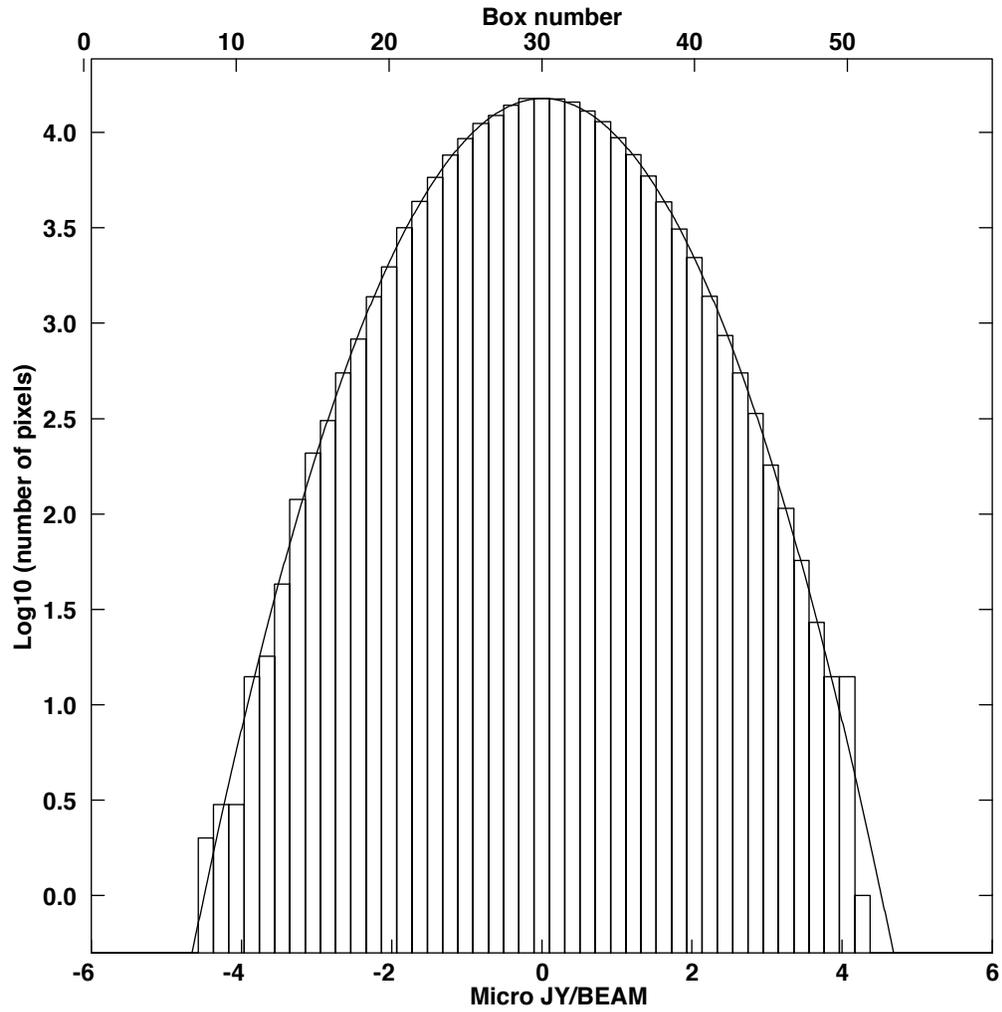}
\caption{Noise distribution.  This noise distribution from one
  source-free region near the edge of our wideband image is a nearly
  perfect Gaussian with rms $\sigma = 1.02\,\mu{\rm Jy~beam}^{-1}$; it
  appears as a parabola on this logarithmic plot.  Abscissa: Log of
  the number of pixels per bin of width $0.2\,\mu{\rm Jy~beam}^{-1}$.
  Ordinate: Peak flux density (microJy/beam).}
\label{noisehistfig}
\end{figure}

\begin{figure}
\epsscale{1.}
\plotone{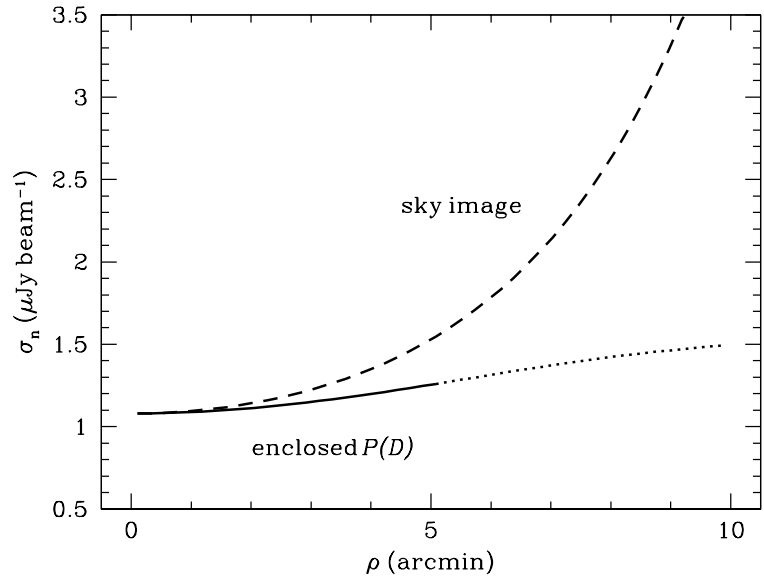}
\vskip -1.0in
\caption{Noise levels in the final image. The rms noise $\sigma_{\rm
    n}$ in the ring of radius $\rho$ in the sky image (dashed curve)
  and in the weighted $P(D)$ distribution of points inside the circle
  of radius $\rho$.  Abscissa: Offset from the pointing center
  (arcmin).  Ordinate: Root-mean-square noise ($\mu{\rm
    Jy\,beam}^{-1}$).}
\label{skynoisefig}
\end{figure}

\begin{figure}
\epsscale{.8}
\plotone{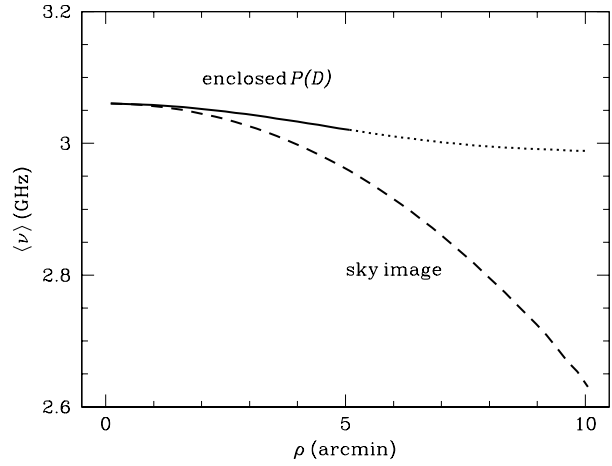}
\vskip -1.in
\caption{Effective frequencies in the final image.  The effective
  frequency $\langle \nu \rangle$ is the frequency at which the flux
  density of a source with spectral index $\alpha = -0.7$ equals the
  local flux density in the ring of radius $\rho$ in the sky image
  (dashed curve) or the average flux density of the weighted $P(D)$
  distribution of points inside the circle of radius $\rho$. Abscissa:
  Offset from the pointing center (arcmin).  Ordinate: Effective
  frequency (GHz).}

\label{skyfreqfig}
\end{figure}

\begin{figure}
\vskip 1.in
\includegraphics[angle=-90, scale=0.7]{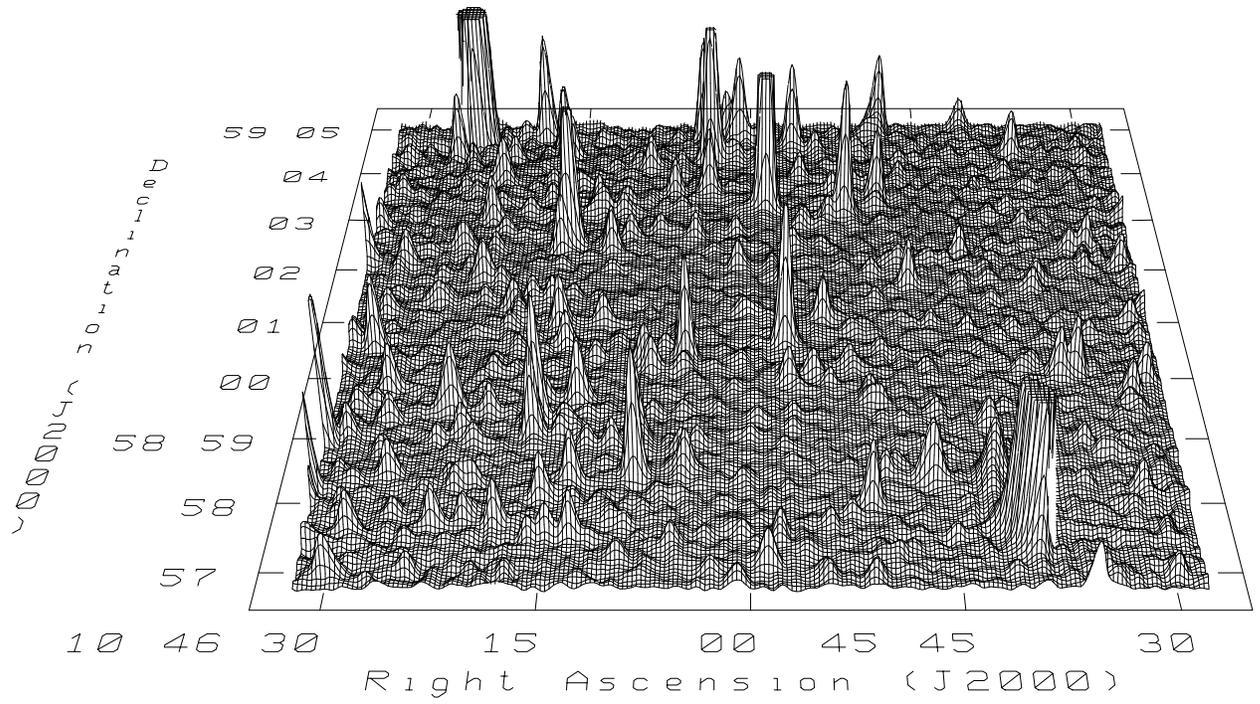}
\caption{Confusion profile.  This profile plot shows the 3 GHz
  confusion amplitude in an 8 arcsec FWHM beam, truncated at 100
  microJy/beam.}
\label{proflfig}
\end{figure}

\begin{figure}
\epsscale{1.}
\plotone{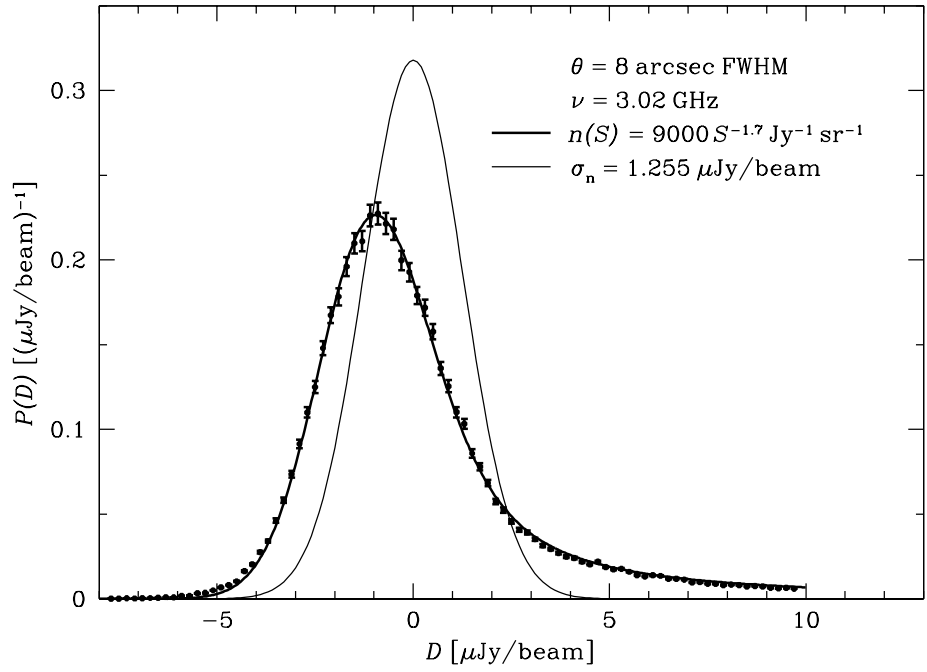}
\vskip -.5in
\caption{Observed $P(D)$ distribution.  The observed probability
  distribution $P(D)$ (data points with error bars) of ``deflections''
  or peak flux densities at 3.02 GHz within a circle of radius $\rho =
  5$ arcmin.  The PSF is a circular Gaussian with FHWM $\theta = 8''$.
  The noise has a Gaussian amplitude distribution with rms
  $\sigma_{\rm n} = 1.255\,\mu{\rm Jy/beam}$ (thin curve).  The thick
  fitted curve is the convolution of the noise distribution with the
  noise-free $P(D)$ distribution \citep{con74} of the power-law source
  distribution $n(S) = 9000 S^{-1.7}{\rm ~Jy}^{-1}{\rm ~sr}^{-1}$.
  Abscissa: Deflection or peak flux density ($\mu{\rm Jy/beam}$) at
  3.02 GHz.  Ordinate: Probability density ($\mu$Jy~beam$^{-1}$).}
\label{pofdbestfitfig}
\end{figure}

\begin{figure}
\epsscale{1.}
\plotone{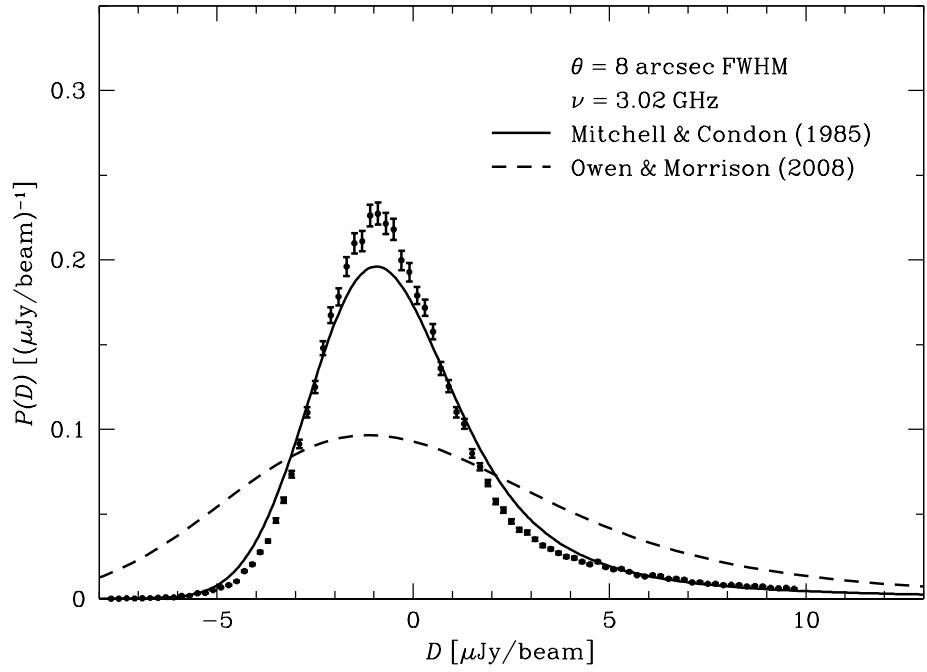}
\vskip -.5in
\caption{Expected $P(D)$ distributions from extrapolations of
  published counts. This figure compares the observed $P(D)$
  distribution (data points with error bars) at 3.02 GHz with
  power-law extrapolations of deep source counts measured at 1.4 GHz,
  assuming a mean spectral index $\langle \alpha \rangle = -0.7$.  The
  continuous curve indicates the expected 3.02 GHz $P(D)$ distribution
  from the power-law extrapolation of the \citet{mit85} 1.4 GHz source
  count $n(S) = 57 S^{-2.2} {\rm ~Jy}^{-1}{\rm ~sr}^{-1}$, and the
  dashed curve corresponds to the \citet{owe08} 1.4~GHz source count
  $n(S) = 6 S^{-2.5} {\rm ~Jy}^{-1}{\rm ~sr}^{-1}$.  Abscissa:
  Deflection or peak flux density ($\mu{\rm Jy/beam}$) at 3.02
  GHz. Ordinate: Probability density ($\mu$Jy~beam$^{-1}$).}

\label{pofdsextrapfig}
\end{figure}

\begin{figure}
\epsscale{1.}
\plotone{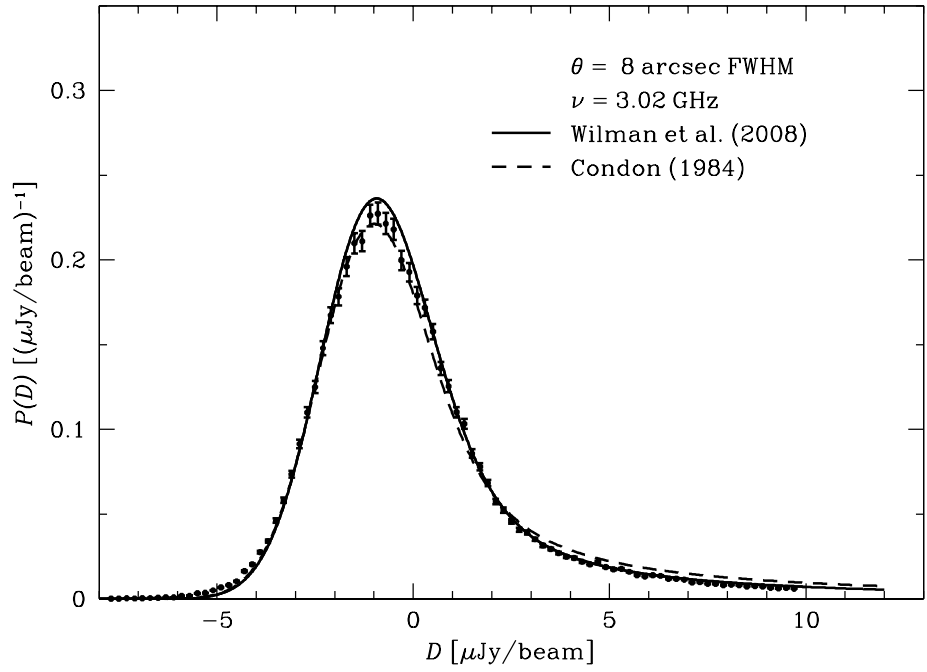}
\vskip -.5in
\caption{Expected $P(D)$ distributions from source-count models. This
  figure compares the observed $P(D)$ (data points with error bars) at
  3.02 GHz with power-law fits to model source counts at 1.4 GHz,
  assuming a mean spectral index $\langle \alpha \rangle = -0.7$.  The
  continuous curve indicates the expected 3.02 GHz $P(D)$ distribution
  for the 1.4 GHz source count $n(S) = 3.45\times10^4 S^{-1.6}$
  predicted by \citet{wil08} and the dashed curve corresponds to $n(S)
  = 1.2\times10^5 S^{-1.5}$ predicted by \citet{con84b}.  Abscissa:
  Deflection or peak flux density ($\mu{\rm Jy/beam}$) at 3.02
  GHz. Ordinate: Probability density ($\mu$Jy~beam$^{-1}$).}

\label{pofdsmodelsfig}
\end{figure}

\begin{figure}
\epsscale{1.}
\plotone{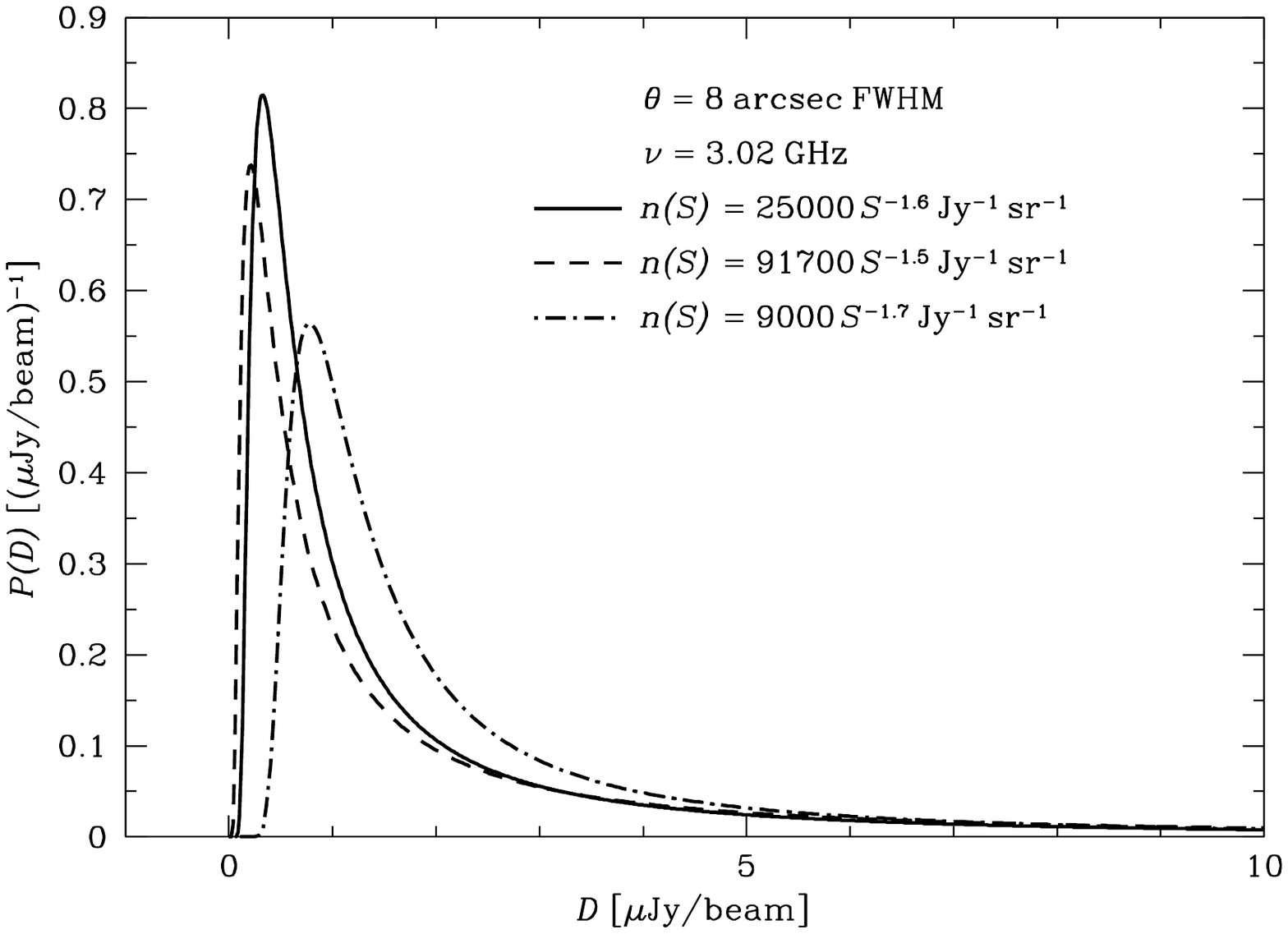}
\vskip -.5in
\caption{Noise-free $P(D)$ distributions. The range of noise-free 3.02
  GHz $P(D)$ distributions in a $FWHM = 8''$ Gaussian beam calculated
  for three power-law approximations to the 3.02 GHz source count that
  are consistent with our data: $n(S) = 25000 S^{-1.6} {\rm~Jy}^{-1}
  {\rm ~sr}^{-1}$ \citep{wil08}, $n(S) = 91700 S^{-1.5} {\rm ~Jy}^{-1}
  {\rm ~sr}^{-1}$ \citep{con84b}, and $n(S) = 9000 S^{-1.7} {\rm
    ~Jy}^{-1} {\rm ~sr}^{-1}$ (this paper).  Abscissa: Deflection or
  peak flux density ($\mu{\rm Jy/beam}$) at 3.02 GHz. Ordinate:
  Probability density ($\mu$Jy~beam$^{-1}$).}

\label{sourcepofd}
\end{figure}

\begin{figure}
\epsscale{1.}
\plotone{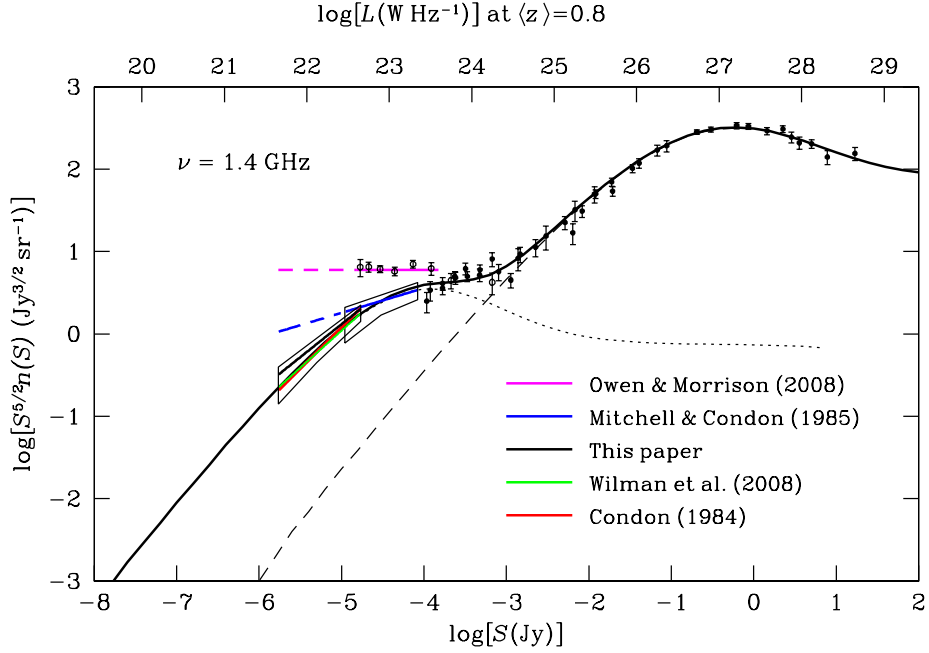}
\vskip -1.in
\caption{Euclidean-normalized source counts.  The plotted
source ource counts at 1.40
  GHz are based on individual sources (data points), the 1.4 GHz $P(D)$
  distribution (blue line and surrounding error box) from
  \citet{mit85}, and our 3.02 GHz $P(D)$ distribution (black line and
  surrounding error box) converted to 1.40 GHz via a mean spectral
  index $\alpha = -0.7$.  The red and green lines are power-law
  approximations to the 1.40 GHz models of \citet{con84b} and
  \citet{wil08}, respectively.  The dashed curve is the contribution
  of AGNs and the dotted curve is the contribution of star-forming
  galaxies, from the \citet{con84b} model. Lower abscissa: log flux
  density (Jy) at 1.40 GHz. Upper abscissa: log spectral luminosity
  (W~Hz$^{-1}$) at 1.4 GHz in the source frame for sources at the
  typical redshift $\langle z \rangle \approx 0.8$.  Ordinate: Source
  count normalized by the source count $n(S) \propto S^{-5/2}$ in a
  static Euclidean universe (Jy$^{3/2}$~sr$^{-1}$).}

\label{eucfig}
\end{figure}

\begin{figure}
\vskip .5in
\epsscale{1.}
\plotone{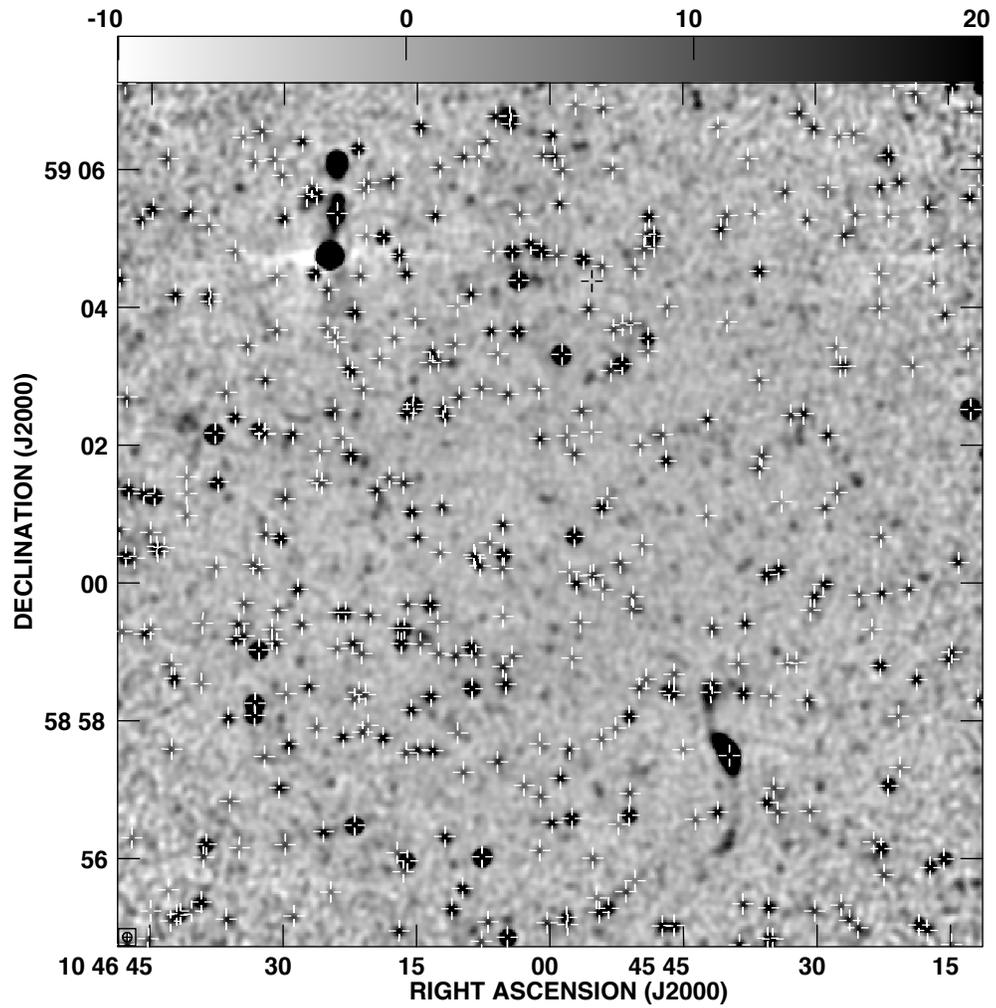}
\caption{Gray-scale sky image at 3~GHz. Sources in the \citet{owe08}
  catalog are indicated by crosses on our gray-scale image. The wedge
  indicates 3 GHz brightnesses between $-10$ and $+20\,\mu{\rm
    Jy\,beam}^{-1}$.}
\label{grayfig}
\end{figure}

\begin{figure}
\vskip .5in
\epsscale{1.}
\plotone{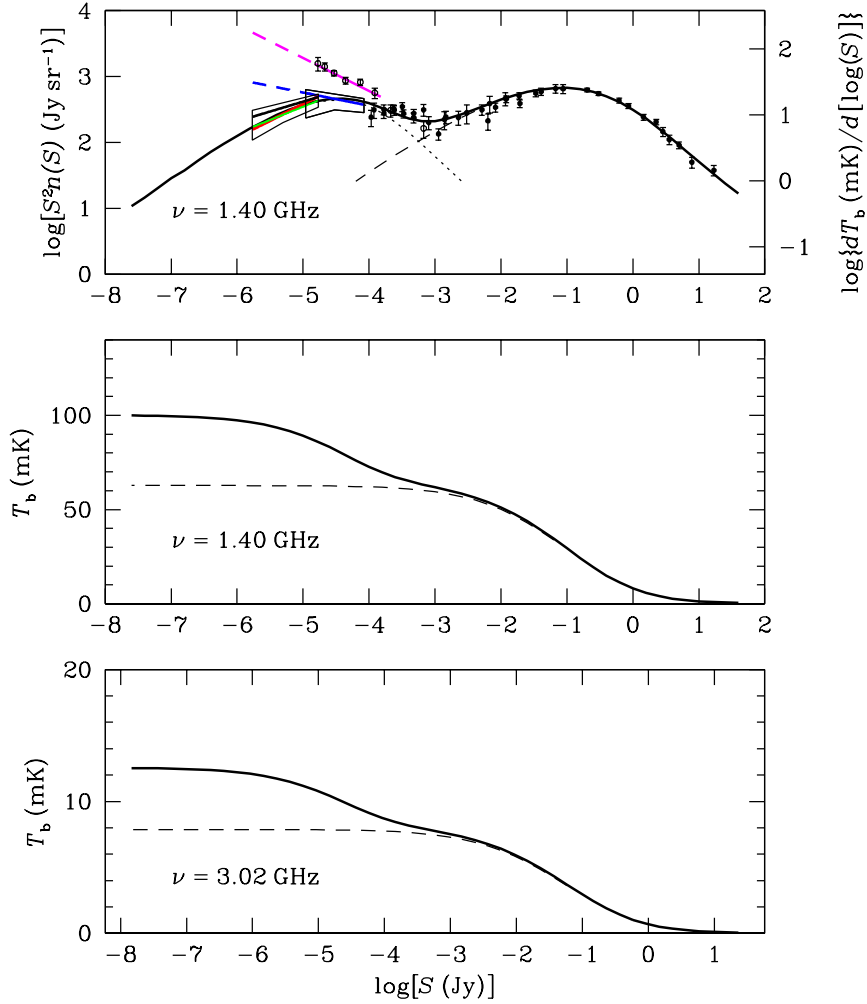}
\caption{Brightness weighted source counts and model sky background
  temperatures at 1.4 and 3.02~GHz.  The top panel shows the weighted
  1.40 GHz source count $S^2 n(S)$, the weighting that reflects the
  contribution of sources to the sky background temperature $T_{\rm
    b}$ per decade of flux density $S$.  The plotted ource counts at
  1.40 GHz are based on individual sources (data points), the 1.4 GHz
  $P(D)$ distribution (blue line and surrounding error box) from
  \citet{mit85}, and our 3.02 GHz $P(D)$ distribution (black line and
  surrounding error box) converted to 1.40 GHz via a mean spectral
  index $\alpha = -0.7$.  The red and green lines are power-law
  approximations to the 1.40 GHz models of \citet{con84b} and
  \citet{wil08}, respectively.  The dashed curve is the contribution
  of AGNs and the dotted curve is the contribution of star-forming
  galaxies, from the \citet{con84b} model.Abscissa: log flux density
  (Jy).  Ordinates, top panel: Log of the 1.40 GHz source count $S^2
  n(S)$ (Jy~sr$^{-1}$) on the left, weighted to be proportional to the
  source contribution $dT_{\rm b} / d[\log(S)]$ (mK) to the sky
  background per decade of flux density on the right.  Ordinate,
  middle panel: Log of the 1.40 GHz sky background temperature (mK)
  contributed by all sources stronger than $S$ at 1.40 GHz.  Ordinate,
  bottom panel: 3.02 GHz sky background temperature (mK) contributed
  by all sources stronger than $S$ at 3.02 GHz.}

\label{tbcountfig}
\end{figure}

\begin{figure}
\vskip .5in
\epsscale{1.}
\plotone{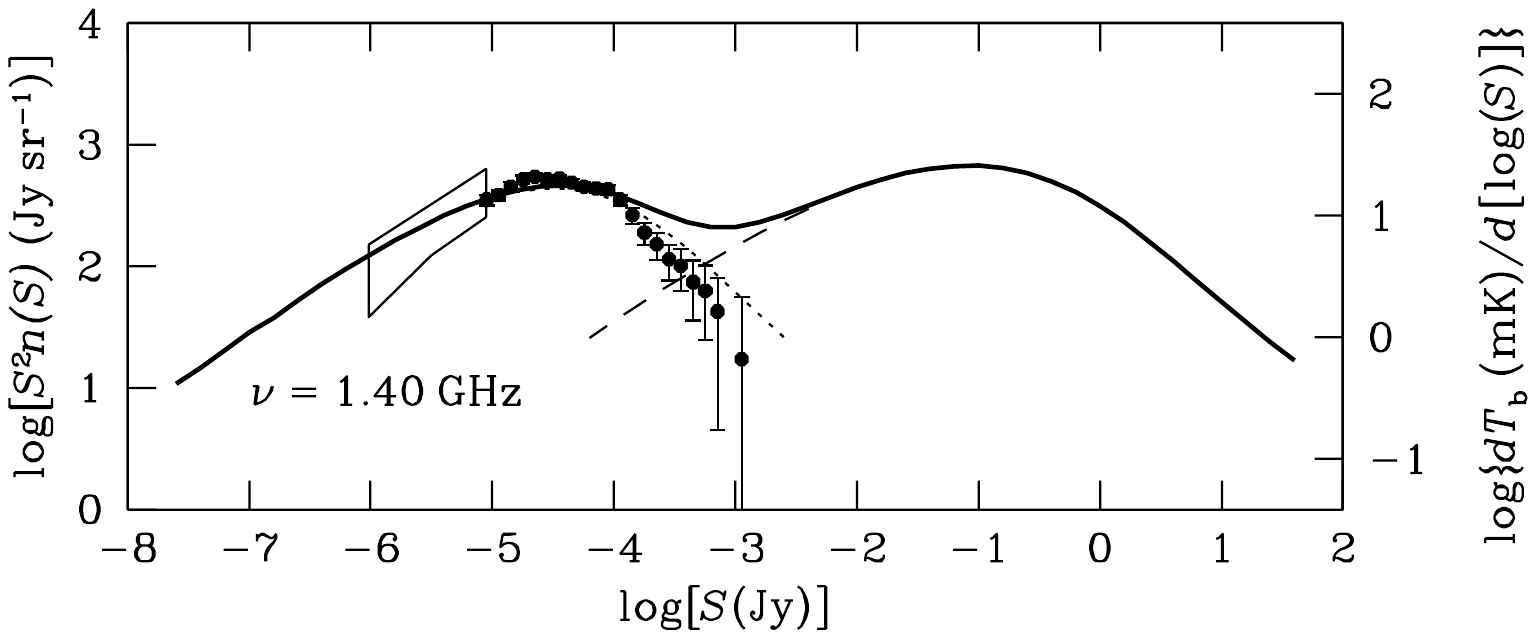}
\vskip -3.5in
\caption{FIR source count converted to 1.4~GHz. The data points and
  $P(D)$ error box show the $\lambda = 160\,\mu$m FIR counts of
  \citet{ber11} converted to 1.4~GHz via the FIR/radio relation
  $\log[S(160\,\mu{\rm m})/S({\rm 1.4~GHz})] = 2.5$.  The curves are
  predictions from the \citet{con84b} 1.4~GHz count model for
  star-forming galaxies (dots), AGNs (dashes), and their sum
  (continuous).  The agreement with the actual 1.4~GHz counts shown in
  Figure~\ref{tbcountfig} is within the errors.  Abscissa: log flux
  density (Jy).  Ordinates: Log of the 1.40 GHz source count $S^2
  n(S)$ (Jy~sr$^{-1}$) on the left, weighted to be proportional to the
  source contribution $dT_{\rm b} / d[\log(S)]$ (mK) to the sky
  background per decade of flux density on the right.}

\label{bertacountfig}
\end{figure}

\begin{figure}
\epsscale{1.}
\plotone{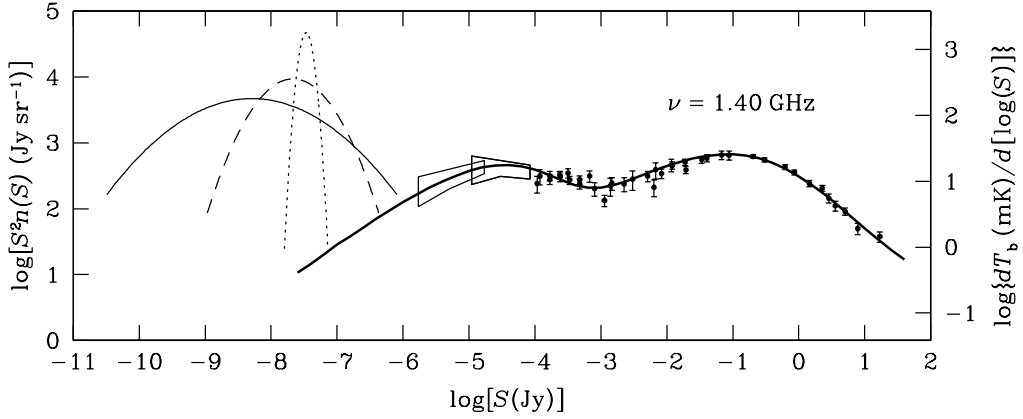}
\vskip -3.in
\caption{Source counts at 1.4~GHz consistent with the ARCADE\,2
  background and our $P(D)$ distribution.  Source counts at 1.40 GHz
  based on individual sources (data points), the 1.4 GHz $P(D)$
  distribution (blue line and surrounding error box) from
  \citet{mit85}, and our 3.02 GHz $P(D)$ distribution (black line and
  surrounding error box) converted to 1.40 GHz via a mean spectral
  index $\alpha = -0.7$.  The red and green lines are power-law
  approximations to the 1.40 GHz models of \citet{con84b} and
  \citet{wil08}, respectively.  The dashed curve is the contribution
  of AGNs and the dotted curve is the contribution of star-forming
  galaxies, from the \citet{con84b} model. The three parabolas at
  nanoJy levels correspond to models for a hypothetical new population
  of radio sources able to produce the ARCADE\,2 excess background
  temperature.  Their widths are $\phi = 0.2$ (dotted), 1.0 (dashed),
  and 2.0 (continuous curve), where $phi$ is defined by
  equation~\ref{gausseq}.  Our upper limit on the rms background
  fluctuation sets upper limits on the flux densities $S_{\rm pk}$ at
  the peaks of these parabolas and hence lower limits on the sky
  density of the new population.  Even in the limit of a very narrow
  flux-density distribution, the number of sources per square arcmin
  is $N > 6\times 10^4$\,arcmin$^{-2}$, which greatly exceeds the $N
  \sim 10^3$\,arcmin$^{-2}$ galaxies brighter than $m = +29$ in the
  Hubble Ultra Deep Field (HUDF).  Abscissa: log 1.4 GHz flux density
  (Jy).  Ordinates: log of the 1.40 GHz source count $S^2 n(S)$
  (Jy~sr$^{-1}$) on the left, weighted to be proportional to the
  source contribution $dT_{\rm b} / d[\log(S)]$ (mK) to the sky
  background per decade of flux density on the right.}
\label{bumpfig}
\end{figure}

\begin{figure}
\epsscale{1.}
\plotone{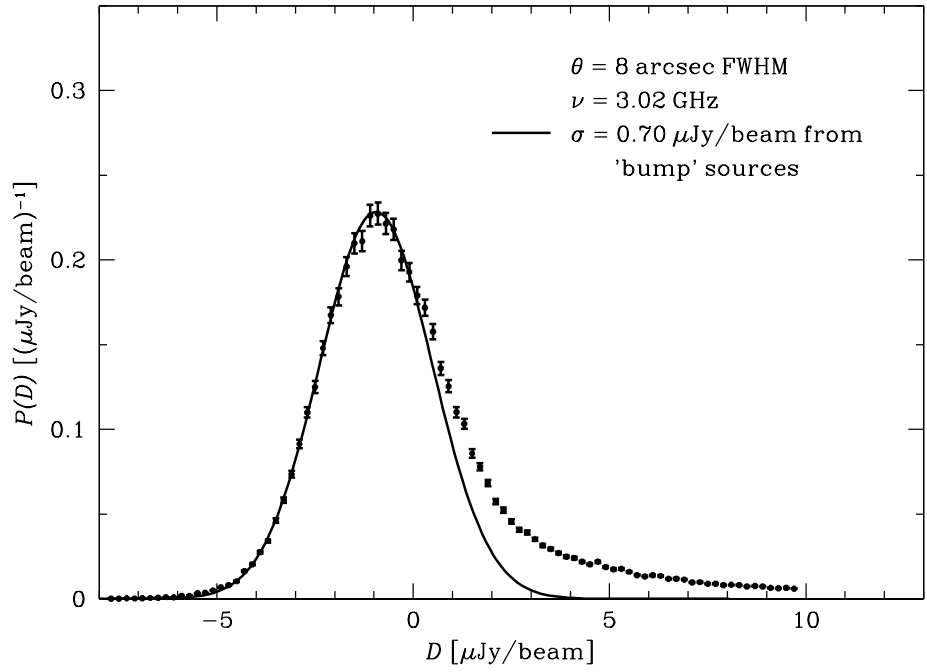}
\vskip -.5in
\caption{The largest possible contribution of numerous faint sources
  to the observed $P(D)$ distribution.  A sharp ``bump'' in the number
  of extremely faint ($S \ll 1\,\mu{\rm Jy}$) faint sources would
  contribute a Gaussian to the observed 3.02 GHz $P(D)$ distribution.
  The rms of that Gaussian cannot exceed $\sigma \approx 0.70\,\mu{\rm
    Jy\,beam}^{-1}$ without excessively broadening our observed $P(D)$
  distribution, no matter how low the source count near $S \sim 1
  \mu{\rm Jy}$.  Abscissa: Deflection or peak flux density ($\mu{\rm
    Jy/beam}$) at 3.02 GHz. Ordinate: Probability density
  ($\mu$Jy~beam$^{-1}$.}

\label{pofdsbump}
\end{figure}

In the noise-limited outer regions of our image, the pixels have a
Gaussian brightness distribution with rms $\sigma_{\rm n} \approx
1~\mu{\rm Jy~beam}^{-1}$. What is the rms fractional uncertainty
$(\Delta\sigma_{\rm n} / \sigma_{\rm n})$ of the rms noise averaged
over a large solid angle $\Omega \gg \Omega_{\rm b}$?  The answer is
complicated because the pixel values are not independent.  Just as the
limited $(u,v)$ plane coverage of data going into our image acts as a
convolving filter to smooth our image of the sky brightness
distribution, it acts as a filter to smooth the image noise
distribution.  The smoothing function for both sky brightness and
noise is the synthesized beam.  Our synthesized beam is closely
approximated by a circular Gaussian (Fig.~\ref{beamproflfig}) with
FWHM $\theta_{\rm s} = 8$ arcsec, so its normalized (peak gain $G =
1$) gain profile can be written
\begin{equation}
G(\theta,\phi) = \exp\biggl[ -4 \ln 2 
\biggl({\theta \over \theta_{\rm s}}\biggr)^2 \biggr]~.
\end{equation}
The corresponding synthesized beam solid angle is
\begin{equation}\label{beamsolidangleeq}
\Omega_{\rm b} \equiv \int G d\Omega =
{\pi \theta_{\rm s}^2 \over 4 \ln 2} \approx 1.13 \theta_{\rm s}^2~.
\end{equation}

It is easy to calculate ($\Delta \sigma_{\rm n}/ \sigma_{\rm n}$)
for a square top-hat beam with FWHM $W$:
\begin{equation}
G(x,y) = 1, ~~-W/2 < x < +W/2,~~-W/2 < y < +W/2
\end{equation}
and $G(x,y) = 0$ elsewhere, so the beam solid angle is $\Omega_{\rm b}
= W^2$. A large area $\Omega$ can be tiled by $N = \Omega /
\Omega_{\rm b} \gg 1$ nonoverlapping beam solid angles.  Averaging
their $N$ independent noise values divides the rms noise by $N^{1/2}$,
so
\begin{equation}\label{tophateq}
{\Delta\sigma_{\rm n} \over \sigma_{\rm n}} = 
\biggl({\Omega_{\rm b} \over \Omega}\biggr)^{1/2} = 
\biggl({1 \over N}\biggr)^{1/2}
\end{equation}
in this case.  This sort of argument seems to underly the common but
incorrect belief that equation~\ref{tophateq} is true for all beamshapes.

The problem is that noise variance $\sigma_{\rm n}^2$, rather than noise
rms $\sigma_{\rm n}$, adds under convolution. The amount of noise smoothing
is proportional to the ``noise beam solid angle''
\begin{equation}
\Omega_{\rm n} \equiv \int G^2 d\Omega
\end{equation}
rather than to the beam solid angle given by equation~\ref{beamsolidangleeq}.
Only in the case of a top-hat beam does $\Omega_{\rm n} = \Omega_{\rm
  b}$.  The square of a Gaussian is also a Gaussian, but its FHWM is
 $\theta_{\rm s} / 2^{1/2}$ so $\Omega_{\rm n} = \Omega_{\rm b} /
2$ for a Gaussian beam.  Thus
\begin{equation}\label{gausseq2}
{\Delta \sigma_{\rm n} \over \sigma_{\rm n}} = 
\biggl({\Omega_{\rm n} \over \Omega}\biggr)^{1/2} =
   \biggl({1 \over 2N}\biggr)^{1/2}~.
\end{equation}

To convince skeptics that this result correctly
describes our data, we selected 31 indepenent source-free
regions from our image, uncorrected for primary attenuation so
$\sigma_{\rm n} \approx 1~\mu{\rm Jy~beam}^{-1}$ in all regions.
Each region covers $100 \times 100 = 10^4$ pixels $\approx 1\,\farcs252$
on a side, so $\Omega_{\rm
  b} \approx 46.3$~pixels and there are $N \approx 216 \gg 1$ beam
solid angles per region.  If equation~\ref{tophateq} were correct for our
Gaussian synthesized beam, we would expect
\begin{equation}
{\Delta \sigma_{\rm n} \over \sigma_{\rm n}} \approx 0.068~.
\end{equation}
Equation~\ref{gausseq2} predicts the lower value
\begin{equation}
{\Delta \sigma_{\rm n} \over \sigma_{\rm n}} \approx 0.048~.
\end{equation}
In the 31 regions the observed mean and its rms scatter are
\begin{equation}
{\Delta \sigma_{\rm n} \over \sigma_{\rm n}} = 0.040 \pm 0.007~.
\end{equation}
This is consistent with equation~\ref{gausseq2} but is four standard deviations
below equation~\ref{tophateq}.

A similar argument applies to calculations of survey speed.  How much
time is needed to make a survey covering a large solid angle $\Omega =
N \Omega_{\rm b}$ using a single Gaussian beam?  If staring time
$\tau$ is needed to reach the desired survey sensitivity at the center
of one beam, the time needed to make the whole survey is $t \approx
2N\tau$, twice the frequently quoted $t \approx N \tau$ \citep{con98}.

Finally, the effective beam area $\Omega_{\rm e}$ for confusion
(Eq.~\ref{omegaeeq}) becomes $\Omega_{\rm e} = \int G^2 d \Omega =
\Omega_{\rm n}$ when $\gamma \rightarrow 3$.  In this limit, the
confusion fluctuations are dominated by the very faintest sources and
the $P(D)$ distribution becomes a Gaussian, just like a noise
distribution.

%% end revision addressing referee's ``second concern''

%% keep appendix from running into references
\hphantom{x}




\clearpage

%% the file, uncomment the \plotone, \plottwo, and
%% If you need a layout that cannot be achieved with \plotone or
%% \plottwo, you can invoke the graphicx package directly with the
%% \includegraphics command or use \plotfiddle. For more information,
%% please see the tutorial on "Using Electronic Art with AASTeX" in the
%% documentation section at the AASTeX Web site,
%% http://www.journals.uchicago.edu/AAS/AASTeX.
%%
%% The examples below also include sample markup for submission of
%% supplemental electronic materials. As always, be sure to check
%% the instructions to authors for the journal you are submitting to
%% for specific submissions guidelines as they vary from

%% This example uses \plotone to include an EPS file scaled to
%% 80% of its natural size with \epsscale. Its caption
%% has been written to indicate that additional figure parts will be
%% available in the electronic journal.

\clearpage

\clearpage

\clearpage

\clearpage

\clearpage

\clearpage

\clearpage

\clearpage

\clearpage

\clearpage

\clearpage

\clearpage

\clearpage

\clearpage

\clearpage

\clearpage

\clearpage

\begin{deluxetable}{cccc}\label{subbandtable}
\tablecaption{Subband Images} \tablewidth{200pt} \tablehead{
  \colhead{Subband} & \colhead{Frequency} & \colhead {S(3C~286)} &
  \colhead{~$\sigma_{\rm n}$} \\ 
\colhead{number} & \colhead {(GHz)} & \colhead{(Jy)} &
  \colhead{($\mu{\rm Jy~beam}^{-1}$)} } \startdata 
01 & 2.0500 & 12.260 & 9.220 \\ 
02 & 2.1780 & 11.877 & {\llap 1}3.190 \\ 
03 & 2.3060 & 11.522 & {\llap 1}8.170 \\ 
04 & 2.4340 & 11.193 & 4.476 \\ 
05 & 2.5620 & 10.885 & 4.244 \\ 
06 & 2.6900 & 10.597 & 4.314 \\ 
07 & 2.8180 & 10.327 & 4.039 \\ 
08 & 2.9460 & 10.073 & 3.719 \\ 
09 & 3.0500 & {\hphantom 1}9.877 & 4.477 \\ 
10 & 3.1780 & {\hphantom 1}9.648 & 3.145 \\ 
11 & 3.3060 & {\hphantom 1}9.432 & 3.067 \\ 
12 & 3.4340 & {\hphantom 1}9.226 & 2.965 \\ 
13 & 3.5620 & {\hphantom 1}9.031 & 2.882 \\ 
14 & 3.6900 & {\hphantom 1}8.845 & 3.468 \\ 
15 & 3.8180 & {\hphantom 1}8.668 & 4.447 \\ 
16 & 3.9460 & {\hphantom 1}8.580 & 4.402 \\ \enddata
\end{deluxetable}

%% The following command ends your manuscript. LaTeX will ignore any text

\end{document}